\documentclass[11pt,a4paper]{article}

\pdfoutput=1

\usepackage{jcappub}
\usepackage{graphicx}
\usepackage{float}
\usepackage{slashed}
\usepackage[font=small, labelfont=bf]{caption}
\usepackage[labelformat=simple]{subcaption}

\usepackage{amsmath, amsthm, amssymb, amsfonts}
\usepackage{multirow}
\usepackage[numbers,sort&compress]{natbib}
\usepackage{hyperref}

\def\beq{\begin{align}}
\def\eeq{\end{align}}

\newcommand{\vo}{\mathcal{V}}

\newcommand{\bi}{\begin{itemize}}
\newcommand{\ei}{\end{itemize}}
\newcommand{\ben}{\begin{enumerate}}
\newcommand{\een}{\end{enumerate}}
\newcommand{\be}{\begin{equation}}
\newcommand{\ee}{\end{equation}}

\newcommand{\F}{\mathcal{F}}
\newcommand{\comments}[1]{}
\def\nn{\nonumber}

\def\SM{{\scriptscriptstyle \rm SM}}

\def\GUT{{\scriptscriptstyle \rm GUT}}

\def\KK{{\scriptscriptstyle \rm KK}}
\def\W{{\scriptscriptstyle \rm W}}

\def\GM{{\scriptscriptstyle \rm GM}}

\def\DT{{\scriptscriptstyle \rm D3}}

\def\vo{{{\cal{V}}}}

\newcommand{\mc}{\mathcal}

\newcommand{\beqa}{\begin{eqnarray}}
\newcommand{\eeqa}{\end{eqnarray}}

\providecommand{\bea}{\begin{eqnarray}}
 
 \providecommand{\rm}{\mathrm}
\providecommand{\eea}{\end{eqnarray}}

\def\K3{{\scriptscriptstyle {\rm K3}}}

\def\F4{{\scriptscriptstyle {\rm F}^4}}

\def\46{{\scriptscriptstyle {\rm 4-6}}}
\def\24{{\scriptscriptstyle {\rm 2-4}}}

\title{Reheating and Dark Radiation after Fibre Inflation}

\author[a,b,c]{Michele Cicoli,}
\author[a]{Gabriel A. Piovano}

\affiliation[a]{\small Dipartimento di Fisica e Astronomia, Universit\`a di Bologna, \\ via Irnerio 46, 40126 Bologna, Italy}
\affiliation[b]{\small INFN, Sezione di Bologna, viale Berti Pichat 6/2, 40127 Bologna, Italy}
\affiliation[c]{\small ICTP, Strada Costiera 11, Trieste 34014, Italy}

\emailAdd{michele.cicoli@unibo.it}
\emailAdd{gabriel.piovano@studio.unibo.it}

\abstract{We study perturbative reheating at the end of fibre inflation where the inflaton is a closed string modulus with a Starobinsky-like potential. We first derive the spectral index $n_s$ and the tensor-to-scalar ratio $r$ as a function of the number of efoldings and the parameter $R$ which controls slow-roll breaking corrections. We then compute the inflaton couplings and decay rates into ultra-light bulk axions and visible sector fields on D7-branes wrapping the inflaton divisor. This leads to a reheating temperature of order $10^{10}$ GeV which requires $52$ efoldings. Ultra-light axions contribute to dark radiation even if $\Delta N_{\rm eff}$ is almost negligible in the generic case where the visible sector D7-stack supports a non-zero gauge flux. If the parameter $R$ is chosen to be small enough, $n_s\simeq 0.965$ is then in perfect agreement with current observations while $r$ turns out to be of order $r\simeq 0.007$. If instead the flux on the inflaton divisor is turned off, $\Delta N_{\rm eff}\lesssim 0.6$ which, when used as a prior for Planck data, requires $n_s\simeq 0.99$. After $R$ is fixed to obtain such a value of $n_s$, primordial gravity waves are larger since $r\simeq 0.01$.}


\begin{document}
\maketitle

\section{Introduction}
\label{Intro}

The expansion rate of the Universe is exponential whenever the first slow-roll parameter $\epsilon\ll 1$. Moreover, if the second slow-roll parameter $\eta \ll 1$, this accelerated expansion can last for enough e-foldings to solve the flatness and horizon problems. However $\eta$ controls the ratio between the inflaton mass and the Hubble scale during inflation, implying that the inflaton has to be a light scalar. However, due to the generic absence of symmetries protecting scalar masses against any kind of quantum corrections, most would-be inflatons would lead to $\eta \sim \mc{O}(1)$. This problem becomes even more severe for large field models with trans-Planckian field ranges since one has to make sure that $\eta\ll 1$ for $\Delta\phi> M_p$.

Thus successful inflationary model building can be achieved only in the presence of a symmetry which can just be postulated from the low-energy effective field theory point of view while it can in principle be derived from a consistent UV framework like string theory \cite{Baumann:2014nda,Burgess:2013sla}. Hence the best case scenario for inflation is a framework where the inflaton is the pseudo Nambu-Goldstone boson of a spontaneously broken effective global symmetry. The inflaton is light since its direction is flat up to small effects which break the symmetry explicitly. The two main Abelian symmetries used in string compactifications for inflationary model building are compact shifts for axions \cite{Pajer:2013fsa} and non-compact rescalings for K\"ahler moduli \cite{Cicoli:2011zz}.

Another crucial issue to trust inflationary models is moduli stabilisation. In fact, one needs to control the directions orthogonal to the inflaton to prevent dangerous runaways and to ensure that the inflationary dynamics is indeed single-field. Moreover, moduli stabilisation is fundamental to fix all the mass and energy scales in the model. 

The presence of a symmetry and full moduli stabilisation are already considered as two considerable achievements for successful inflationary model building and most of the models in the literature, which are considered to be good working examples, satisfy these two requirements. However in order to check the consistency of each inflationary model, one has to study additional formal and phenomenological features. From the theoretical point of view, one has to provide a concrete Calabi-Yau embedding with an explicit orientifold involution and brane setup which are compatible with tadpole cancellation and the presence of a chiral SM-like visible sector \cite{Cicoli:2016xae, Cicoli:2017axo, Cicoli:2017shd}. On the other hand, more phenomenological implications of any model, like the predictions for cosmological observables as the scalar spectral index $n_s$ or the tensor-to-scalar ratio $r$, depend crucially on reheating and the details of the post-inflationary evolution of our Universe. In fact, one can extract trustable predictions from an inflationary model only by developing a thorough understanding of what happens after the end of inflation since the exact number of efoldings $N_e$, corresponding to horizon exit of CMB modes, depends on the reheating temperature $T_{\rm rh}$ and the fact that the Universe is radiation dominated or not all the way from reheating to Big Bang Nucleosynthesis (BBN). Notice that non-standard post-inflationary cosmologies tend to arise rather naturally from string theory due to the presence of long-lived moduli which can give rise to periods of matter domination before BBN \cite{Acharya:2008bk,Dutta:2014tya,Cicoli:2016olq}. Given that $N_e$ determines the point where $n_s$ and $r$ have to be evaluated, one cannot make a clear prediction without studying reheating. 

Moreover, $n_s$ is a quantity which should match the most recent value from cosmological data, while the value of $r$ turns out to be an actual prediction for future observations. However the spectral index inferred from CMB data depends crucially on the choice of some fundamental parameters. A crucial quantity which is positively correlated with $n_s$ is the effective number of neutrino-like species $N_{\rm eff}$ \cite{Ade:2015xua}. If this quantity is set equal to its SM value, $\Delta N_{\rm eff}\equiv N_{\rm eff}- 3.046=0$, Planck data give $n_s=0.965\pm 0.004$ ($68\%$) \cite{Aghanim:2018eyx}. On the contrary, if one sets as a prior $\Delta N_{\rm eff}=0.39$, Planck data lead to a considerably larger spectral index, $n_s=0.983\pm 0.006$ ($68\%$) \cite{Ade:2015xua}. Hence it is crucial to predict the precise value of $\Delta N_{\rm eff}$ in order to know which value of $n_s$ should be matched. 

Interestingly, another rather generic feature of 4D string models is the presence of ultra-light axions which tend to be the supersymmetric partners of K\"ahler moduli \cite{Svrcek:2006yi, Conlon:2006tq, Arvanitaki:2009fg, Cicoli:2012sz}. These fields can be produced either directly from the inflaton decay or from the decay of the lightest modulus which triggers reheating \cite{Cicoli:2012aq, Higaki:2012ar, Angus:2014bia, Hebecker:2014gka, Cicoli:2015bpq,Acharya:2015zfk}. Given that the moduli are SM-singlets, in general the branching ratio for their decay into ultra-light axions is non-negligible, leading to a non-zero contribution to $\Delta N_{\rm eff}$. Notice that the mass of these axions tends to be exponentially smaller than the gravitino mass, and so these pseudo-scalars are so light that they never become non-relativistic. The goal of each string inflationary model which aims at yielding trustable predictions is thus to provide a detailed description of reheating which can allow to compute $T_{\rm rh}$ and $\Delta N_{\rm eff}$, and to follow all the post-inflationary evolution all the way down to BBN. Perturbative reheating after the end of some string inflationary models has already been studied in \cite{Cicoli:2010ha,Cicoli:2010yj} without however taking into account the crucial issue of dark radiation overproduction due to a non-zero inflaton branching ratio into ultra-light axions. 

In this paper we shall study perturbative reheating for a well-known class of string models called \emph{Fibre Inflation} \cite{Cicoli:2008gp, Burgess:2016owb}. These constructions are within the framework of type IIB LVS models \cite{Balasubramanian:2005zx, Cicoli:2008va}. The underlying Calabi-Yau manifold has a fibred structure and its volume is controlled by two K\"ahler moduli associated with the volumes of the 2D $\mathbb{P}^1$ base and the 4D K3 or $T^4$ fibre \cite{Cicoli:2011it}. The internal space features also additional rigid del Pezzo divisors which support non-perturbative effects. At leading order in an expansion in inverse powers of the exponentially large (in string units) internal volume, $\mc{O}(\alpha'^3)$ contributions to the scalar potential \cite{Becker:2002nn} and non-perturbative effects \cite{Blumenhagen:2009qh} fix all the moduli expect for three directions which remain flat. These correspond to the K\"ahler modulus $\tau_1$ parameterising the fibre volume and two axions $\theta_1$ and $\theta_2$ associated respectively with the reduction of the 10D bulk 4-form $C_4$ on the fibre divisor and the 4-cycle containing the 2D base. At perturbative level, only $\tau_1$ develops a potential by either 1-loop open string effects \cite{Berg:2005ja,Berg:2007wt,Cicoli:2007xp} or higher derivative $\alpha'$ corrections to the effective action \cite{Ciupke:2015msa, Grimm:2017okk}. Depending on which of these contributions to the inflationary potential is present or dominates (due to topological properties, details of the brane setup and tuning of flux-dependent coefficients), several slightly different inflationary models can emerge \cite{Cicoli:2008gp,Broy:2015zba,Cicoli:2016chb}. However all of these models are characterised by a similar shape of the inflaton potential which resembles Starobinsky inflation \cite{Starobinsky:1980te} and supergravity $\alpha$-attractors \cite{Kallosh:2013maa, Kallosh:2017wku} since it features a constant plus negative exponentials which generate a trans-Planckian inflationary plateau. The inflaton range is around $\Delta\phi\sim\mc{O}(5)\, M_p$ with larger values bounded by the size of the underlying K\"ahler cone \cite{Cicoli:2018tcq}. Hence the final prediction for primordial gravity waves is $0.005\lesssim r\lesssim 0.01$. The stability of the inflationary potential against further corrections is guaranteed by an effective rescaling shift symmetry \cite{Burgess:2014tja} and globally consistent Calabi-Yau embeddings with consistent brane setup and a chiral visible sector have been recently constructed in \cite{Cicoli:2016xae, Cicoli:2017axo}. 

The actual predictions for $n_s$ and $r$ depend just on two quantities: the number of efoldings $N_e$ between CMB horizon exit and the end of inflation, and the naturally small coefficient $R \sim g_s^4 \ll 1$ (for a string coupling $g_s\lesssim 0.1$ in the perturbative regime) controlling the strength of the positive exponentials which dominate the inflationary potential at large field values and destroy its flatness. Thus the first step of our analysis will be the derivation of the relations $n_s=n_s(N_e, R)$ and $r=r(N_e, R)$. In turn, $N_e$ depends on the reheating temperature $T_{\rm rh}$ and the equation of state parameter $w_{\rm rh}$ for the reheating epoch. In this case, the inflaton $\tau_1$ oscillates around the minimum after the end of inflation and behaves as a classical condensate which redshifts as matter, setting $w_{\rm rh}=0$, before decaying into hidden and visible sector particles. 

In this paper, we shall focus only on the perturbative decay of the inflaton since studies of oscillon formation in Fibre Inflation have revealed that there is no efficient particle production due to non-perturbative preheating effects after the end of inflation \cite{Antusch:2017flz}. However ref. \cite{Gu:2018akj} has recently claimed that parametric self-resonance can produce inflaton quanta after the end of Fibre Inflation even it is not efficient enough to convert all the energy of the inflaton condensate into particle production. The disagreement between the two results might be due to the fact that the analysis of \cite{Antusch:2017flz} is based on a numerical study while the analysis of \cite{Gu:2018akj} is based on an analytical approximation. Due to the intrinsically non-linear nature of the preheating phase, we think that numerical results are more trustable. Hence we would tend to conclude that the process through which the inflaton energy gets transferred into SM degrees of freedom is dominated by a purely perturbative dynamics. Let us mention that, even if there is production of inflaton particles and/or the formation of oscillons after the end of inflation, one would still need to study their perturbative decay unless oscillons collapse into black holes before decaying (see \cite{Helfer:2016ljl} for the study of the stability of oscillon solutions for axion-like potentials). 

Let us stress also that, given that the inflaton is the lightest K\"ahler modulus, no period of matter domination can arise between reheating from the inflaton decay and BBN. The two axions $\theta_1$ and $\theta_2$ could in principle dominate the energy density after the end of inflation but they receive a potential only via tiny non-perturbative corrections to the superpotential which give them a mass that is exponentially smaller than the gravitino mass. Hence these pseudo-scalars are almost massless and could just behave as relativistic degrees of freedom which belong to the hidden sector and can contribute to $\Delta N_{\rm eff}$. Present upper bounds on this quantity set strong constraints on our brane setup. In fact, if the SM lives on D3-branes at singularities \cite{Conlon:2008wa,Cicoli:2012vw}, the visible sector is sequestered from the bulk \cite{Blumenhagen:2009gk, Aparicio:2014wxa}, resulting in an effective decoupling of the inflaton from the the visible sector degrees of freedom \cite{Cicoli:2012aq, Higaki:2012ar}. Consequently, the main inflaton decay channels are just to $\theta_1$ and $\theta_2$ leading to a value of $\Delta N_{\rm eff}$ which is way too large \cite{Angus:2014bia}. Therefore the SM is forced to live on stacks of D7-branes wrapped around the inflaton cycle. In this case the visible sector is unsequestered and the soft terms turn out to be of order the gravitino mass \cite{Conlon:2006wz}. Given that the inflaton is much lighter than the gravitino, the inflaton decay to supersymmetric particles is kinematically forbidden and the dominant inflaton decay channels are into Higgses, SM gauge bosons and hidden sector ultra-light bulk axions \cite{Hebecker:2014gka}. Thus in this case the branching ratio of the inflaton decay into $\theta_1$ and $\theta_2$ is much smaller than unity, resulting in dark radiation in agreement with observational bounds. 

We will find two separate cases depending on the gauge flux on the visible sector stack of D7-branes wrapping the fibre divisor. In the generic case where this flux is non-zero, the inflaton coupling to visible sector gauge bosons is enhanced. Hence $\Delta N_{\rm eff}$ turns out to be negligible, in perfect agreement with current observations since Planck+galaxy BAO data give $N_{\rm eff} = 2.99\pm 0.17$ ($68\%$) \cite{Aghanim:2018eyx}. On the other hand, when the gauge flux is absent, ultra-light axions generate a considerable contribution to dark radiation of order $\Delta N_{\rm eff}\lesssim 0.6$. Notice that such a large value of $\Delta N_{\rm eff}$ is not necessarily ruled out since a combined Planck+galaxy BAO+HST analysis gives $N_{\rm eff} = 3.27 \pm 0.15$ ($68\%$) \cite{Aghanim:2018eyx}. Moreover, when LyaF BAO measurements are added to Planck+galaxy BAO data, the amount of dark radiation increases to $N_{\rm eff} = 3.43 \pm 0.26$ ($68\%$) \cite{Aubourg:2014yra}. Finally in a comprehensive combination of SN and LyaF BAO measurements with Planck+galaxy BAO+HST data, $N_{\rm eff}$ is pulled to similarly high values of order $N_{\rm eff} = 3.41 \pm 0.22$ ($68\%$) \cite{Riess:2016jrr}.

Our study of reheating will also allow us to fix the reheating temperature at $T_{\rm rh} \simeq 10^{10}$ GeV which implies $N_e\simeq 52$. Notice that a similar result for the reheating temperature of Fibre Inflation models has already been found in \cite{Cicoli:2010ha} by considering just the inflaton decay into visible sector gauge bosons. Our analysis is however deeper since it includes also the study of the additional leading order inflaton decay channels into Higgses and ultra-light bulk axion and the corresponding model building constraints from the requirement of avoiding the overproduction of axionic dark radiation. The two results for the order of magnitude of the reheating temperature however agree since we will find that the inflaton decay into gauge bosons tends to dominate over the decay into Higgs degrees of freedom. We shall then plug $N_e\simeq 52$ in the inflationary relations $n_s=n_s(N_e, R)$ and $r=r(N_e, R)$ which simplify to $n_s=n_s(R)$ and $r=r(R)$. The microscopic parameter $R$ can be fixed from $n_s=n_s(R)$ by using the result of our computation of $\Delta N_{\rm eff}$ as a precise prior for Planck data. Once $R$ is fixed, the relation $r=r(R)$ gives the prediction of Fibre Inflation models for the amplitude of primordial gravity waves. Notice that our analysis is complementary to the study of reheating for Fibre Inflation performed in \cite{Cabella:2017zsa} which worked out $T_{\rm rh}$ as a function of $n_s$ and $w_{\rm rh}$ independently of the microscopic details of the model.

In the generic case where $\Delta N_{\rm eff}$ is close to zero, Planck data give $n_s$ centered around $n_s\simeq 0.965$ \cite{Aghanim:2018eyx} which can be easily obtained for a sufficiently small value of the microscopic parameter $R$ of order $R\ll R_0\equiv 2.7\cdot 10^{-5}$. The tensor-to-scalar ratio then turns out to be of order $r\simeq 0.007$. If instead $\Delta N_{\rm eff}\lesssim 0.6$ since the gauge flux on visible D7-stack is vanishing, we shall consider $n_s\simeq 0.99$ as the number to be matched to fix $R$ from $n_s=n_s(R)$. In fact, Fig. 21 of \cite{Ade:2015xua} (see also Fig. 20 of the same paper for $n_s$ values in deviations from the $\Lambda$CDM model with $\Delta N_{\rm eff}\neq 0$) gives a spectral index in the range $n_s \simeq 0.98 - 0.99$ at $1\sigma$ for $\Delta N_{\rm eff}=0.39$, and so it seems reasonable to consider $n_s\simeq 0.99$ for $\Delta N_{\rm eff}\lesssim 0.6$. In this case $R \gtrsim R_0$ and the relation $r=r(R)$ leads to larger primordial tensor modes since $r \simeq 0.01$. 

Notice that in both cases, due to the rather large value of $r$, the inflationary scale is high, $M_{\rm inf}\simeq 10^{16}$ GeV, and the gravitino mass is around $m_{3/2}\simeq 5\cdot 10^{15}$ GeV. An interesting implication of this result is that the visible sector cannot be a simple constrained MSSM with universal scalar and gaugino masses since the scale of supersymmetry breaking would be too high to allow for a correct Higgs mass around $125$ GeV \cite{Bagnaschi:2014rsa}. Therefore in Fibre Inflation models the visible sector has to have a richer structure, involving non-universal soft terms \cite{Ellis:2017erg}, additional fields like in the NMSSM \cite{Zarate:2016jch} or a larger gauge group.        

The plan of this paper is as follows. In Sec. \ref{SecInfl} we briefly describe the dynamics of Fibre Inflation models and we derive the dependence of the spectral index and the tensor-to-scalar ratio on the number of efoldings and the coefficient of the corrections which steepen the potential at large field values. In Sec. \ref{SecReheat} we first illustrate the brane setup and derive the inflaton couplings to all visible sector particles and bulk axions, and then compute the reheating temperature, the number of efoldings, the amount of dark radiation and the final predictions for $n_s$ and  $r$. We finally discuss our results and present our conclusions in Sec. \ref{SecConcl}.

\section{Fibre inflation}
\label{SecInfl}

\subsection{General idea}

Fibre Inflation \cite{Cicoli:2008gp,Burgess:2016owb} is a class of type IIB string inflationary models where the inflaton is a K\"ahler modulus which behaves as the pseudo Nambu-Goldstone boson of a slightly broken non-compact Abelian symmetry \cite{Burgess:2014tja}. The perturbative K\"ahler potential of type IIB compactifications looks like:
\be
K = K_{\rm tree} + K_{\alpha'} + K_{g_s} \,,
\ee
where the tree-level part is: 
\be
K_{\rm tree}=-2\ln\vo\,,
\ee 
and leads to the famous no-scale cancellation. Hence at tree-level the scalar potential for all $h^{1,1}$ K\"ahler moduli is flat. On the other hand, the leading order $\alpha'$ \cite{Becker:2002nn} and $g_s$ corrections to $K$ \cite{Berg:2005ja,Berg:2007wt,Cicoli:2007xp} in 4D Einstein frame scale as:
\be
K_{\alpha'} = -\frac{c_{\alpha'}}{g_s^{3/2} \vo} \qquad\text{and}\qquad K_{g_s}= g_s  \sum_i \frac{c_{g_s}^i t_i}{\vo}\,,
\ee
where the $t$'s are 2-cycle volumes while $c_{\alpha'}$ and $c_{g_s}^i$ are $\mc{O}(1)$ coefficients. Focusing on just one K\"ahler modulus, the ratio between $\alpha'$ and $g_s$ effects scales as:
\be
\frac{K_{\alpha'}}{K_{g_s}} = \frac{c_{\alpha'}}{c_{g_s}} \frac{1}{g_s^{5/2} \vo^{1/3}}\ll 1 \qquad\text{for}\,\,\, \vo\gg 1\,.
\ee
However due to the extended no-scale structure \cite{Cicoli:2007xp}, there is a cancellation in the leading contribution of $K_{g_s}$ to the scalar potential, implying that $g_s$ effects are subdominant with respect to $\alpha'$ effects at the level of the scalar potential even if they are the leading perturbative contributions to $K$. In fact $\alpha'$ and $g_s$ corrections to the scalar potential scale as:
\be
V_{\alpha'} = \frac{c_{\alpha'} W_0^2}{g_s^{3/2} \vo^3}
\qquad\text{and}\qquad V_{g_s} = g_s^2 \sum_i c_{g_s}^i \left(\partial^2_{\tau_i}K_{\rm tree}\right) \frac{W_0^2}{\vo^2}\,,
\label{Vgs}
\ee
where $\tau_i$ denotes 4-cycle moduli and $W_0$ the tree-level flux superpotential. In the case of just one K\"ahler modulus $\vo\sim \tau^{3/2}$, and so $\partial^2_{\tau_i}K_{\rm tree} \sim \vo^{-4/3}$, implying that: 
\be
\frac{V_{\alpha'}}{V_{g_s}} \sim \frac{\vo^{1/3}}{g_s^{7/2}} \gg 1 \qquad\text{for}\,\,\, \vo\gg 1\,\,\text{and}\,\,\,g_s\lesssim 0.1\,.
\ee
Therefore the flatness of the tree-level potential is broken at leading order by $\alpha'$ corrections which however depend just on the overall volume $\vo$, and so lift just one direction in K\"ahler moduli space. This means that all $(h^{1,1}-1)$ directions orthogonal to the overall volume remain exactly flat at leading order, and so are very promising inflaton candidates. These remaining flat directions can be lifted by subdominant perturbative $g_s$ effects which can depend on all K\"ahler moduli, or even by higher derivative $\alpha'$ corrections to the effective action \cite{Ciupke:2015msa, Grimm:2017okk}. 

Moreover, these $(h^{1,1}-1)$ directions orthogonal to $\vo$ enjoy an effective approximate shift symmetry since both $K_{\rm tree}$ and $K_{\alpha'}$ depend just on $\vo$ \cite{Burgess:2014tja}. This shift symmetry is broken by $g_s$ effects since $K_{g_s}$ depends on all K\"ahler moduli but the fact that this is a small effect with respect to the tree-level contribution guarantees that higher dimensional operators are sufficiently suppressed. In fact, these operators are suppressed by both powers of $M_p$ and the small ratio $K_{g_s}/K_{\rm tree}\sim g_s\vo^{-2/3}(\ln \vo)^{-1}\ll 1$.

\subsection{Inflationary potential}

\subsubsection*{Compactification manifold}

Fibre Inflation models involve a Calabi-Yau manifold with at least 3 K\"ahler moduli \cite{Cicoli:2008gp}:

\ben
\item $T_1=\tau_1 +{\rm i}\,\theta_1$: $\tau_1$ plays the r\^ole of the inflaton and parametrises the volume of a K3 or $T^4$ divisor $D_1$ fibred over a $\mathbb{P}^1$ base. This field is stabilised at subleading order due to string loop corrections to $K$ \cite{Cicoli:2008va} or higher derivative $F^4$ terms \cite{Ciupke:2015msa}. The field $\theta_1$ comes from the reduction of the 10D bulk form $C_4$ over $D_1$ and it is fixed due to tiny non-perturbative corrections to the superpotential $W$ \cite{Cicoli:2017zbx,Cicoli:2018kdo}. This field is much lighter than the inflaton and acquires isocurvature fluctuations during inflation. However present strong bounds on isocurvature fluctuations \cite{Akrami:2018odb} due not apply to our case since $\theta_1$ is too light to behave as dark matter.

\item $T_2= \tau_2+ {\rm i}\,\theta_2$: $\tau_2$ parametrises the volume of the base $D_2$ of the fibration and mainly controls the overall volume $\vo$ which is stabilised at leading order via $\alpha'$ corrections. The volume modulus $\vo$ is a spectator during inflation since it turns out to be heavier than the inflaton. The axion $\theta_2$ comes from the reduction of $C_4$ over $D_2$ and, similarly to $\theta_1$, is fixed due to tiny non-perturbative effects and it is much lighter than the inflaton during inflation.

\item $T_3= \tau_3+ {\rm i}\,\theta_3$: $\tau_3$ controls the size of a blow-up mode $D_3$ which is required to perform a full stabilisation of $\vo$ at leading order. Both $\tau_3$ and $\theta_3$ are fixed by non-perturbative corrections to $W$ and are heavier than $\tau_1$ during inflation. 
\een
Expanding the K\"ahler form $J$ in a basis of dual 2-forms as:
\be
J= t_1 \hat{D}_1 +t_2 \hat{D}_2-t_3 \hat{D}_3\,,
\ee
the Calabi-Yau volume can expressed in terms of 2-cycle moduli as \cite{Cicoli:2011it}:
\be
\vo= \frac16\,\int_{{\rm CY}} J\wedge J\wedge J =\frac16\left(
3 k_{122}\, t_1 t_2^2 - k_{333}\, t_3^3\right),
\label{volts}
\ee
where $t_1$ is the volume of the $\mathbb{P}^1$ base while $t_2^2$ is the size of the K3 or $T^4$ fibre. 
Using the expressions of the 4-cycle moduli:
\be
\tau_1 =\frac{1}{2}\int_{{\rm CY}} J\wedge J\wedge \hat{D}_1 = \frac12\,k_{122}t_2^2,
\qquad \tau_2 = k_{122}t_1 t_2\,,\qquad \tau_3 = \frac{1}{2}k_{333}t_3^2\,,
\ee
the volume (\ref{volts}) can be rewritten as:
\be
\vo\,=\,\alpha \left( \sqrt{\tau_1} \tau_2-\gamma \tau_3^{3/2}\right), 
\label{initialVolume}
\ee
where $\alpha=1/\sqrt{2 k_{122}}$ and $\gamma = \frac 23 \sqrt{k_{122}/k_{333}}$.

\subsubsection*{Effective action}

The 4D $N=1$ supergravity theory is characterised by the following K\"ahler potential $K$ and superpotential $W$. The tree-level $K$ together with the leading $\alpha'$ and $g_s$ perturbative corrections reads: 
\be 
K = K_{\rm tree}+ K_{\alpha'} + K_{g_s}= -2\ln\vo - \frac{\xi}{g_s^{3/2}\vo} + K_{g_s}^\KK + K_{g_s}^\W\,,
\label{K0pot}
\ee
with $\xi =-\frac{\zeta(3)\,\chi({\rm CY})}{2 \,(2 \pi)^3}$ where $\chi({\rm CY})$ is the Calabi-Yau Euler number \cite{Becker:2002nn}. The 1-loop open string correction $K_{g_s}^\KK$ comes from the tree-level propagation of closed Kaluza-Klein strings between non-intersecting stacks of branes, while $K_{g_s}^\W$ is due to the tree-level exchange of winding strings between intersecting branes \cite{Berg:2005ja}. These corrections have been conjectured to take the form \cite{Berg:2007wt}:
\be
K_{g_s}^\KK = g_s \sum_i  \frac{c_i^\KK t_i^\perp}{\vo}\,\qquad\text{and}\qquad K_{g_s}^\W = \sum_i \frac{c_i^\W}{t_i^\cap \vo}\,,
\label{Kgs}
\ee
where $t_i^\perp$ is the 2-cycle perpendicular to the $i$-th couple of parallel branes, while $t_i^\cap$ is the 2-cycle which is the intersection locus of the $i$-th couple of intersecting D7-branes/O7-planes. Notice that these loop corrections have been computed explicitly only for toroidal orientifolds \cite{Berg:2007wt}. However in our case (\ref{Kgs}) is a very reasonable conjecture for several reasons: ($i$) for inflationary purposes we need to infer only the dependence of the K\"ahler moduli, and this can be easily deduced from writing the Kaluza-Klein and the winding mass scales in terms of the $T$-moduli and from the Weyl rescaling to go to 4D Einstein frame \cite{Berg:2007wt}; ($ii$) the form of the scalar potential following from (\ref{Kgs}) has been shown to reproduce the exact volume scaling of the Coleman-Weinberg potential \cite{Cicoli:2007xp}; ($iii$) the volume form (\ref{volts}) $\vo \sim t_1 t_2^2$ has basically the same structure of the toroidal case $\vo \sim t_1 t_2 t_3$ after identifying $t_3$ with $t_2$. 

The superpotential $W$ receives a tree-level contribution from background fluxes which is just a constant $W_0$ after dilaton and complex structure moduli stabilisation. Moreover $W$ can depend on all K\"ahler moduli at non-perturbative level:
\be 
W = W_0  + A_1 \,e^{- a_1 T_1} + A_2\, e^{-a_2 T_2} + A_3 \,e^{- a_3 T_3}\,.
\label{W0pot}
\ee
The parameters $A_i$ and $a_i$ are constants. In particular, $a_i = 2\pi/N_i$ with $N_i=1$ in the case of an ED3-instanton while $N_i$ is the rank of the condensing gauge group in the case of gaugino condensation on D7-branes \cite{Blumenhagen:2009qh}.

The F-term scalar potential originating from (\ref{K0pot}) and (\ref{W0pot}), can be organised in an expansion in inverse powers of $\vo\gg 1$:
\be
V= V_{\mc{O}(\vo^{-3})}(\vo,\tau_3,\theta_3) +V_{\mc{O}(\vo^{-10/3})}(\tau_1)+V_{\mc{O}(\vo^{-4/3} e^{-\vo^{2/3}})}(\theta_1,\theta_2)\,,
\label{Pr}
\ee
where we did not include the tree-level scalar potential which scales as $\vo^{-2}$ since the no-scale cancellation implies that $V$ is exactly flat for all $T$-moduli at this order of approximation. The potential beyond tree-level can be studied order by order in the inverse volume expansion, showing that at each order it can provide stable minima for some of the moduli. Notice that in each contribution in (\ref{Pr}) we have shown explicitly only the dependence on the moduli which are fixed at that order of approximation. 

\subsubsection*{Moduli stabilisation at $\mc{O}(\vo^{-3})$}

Focusing on the large volume limit $\sqrt{\tau_1} \tau_2 \gg \tau_3^{3/2}$,\footnote{Without loss of generality, from now on we shall set $\gamma  = 1$, $W_0=|W_0|$ and $A_i=|A_i|$ $\forall i$.} the leading contributions to the potential for the $T$-moduli come from $\alpha'$ corrections to $K$ and $T_3$-dependent non-perturbative effects which give a typical LVS potential \cite{Balasubramanian:2005zx,Cicoli:2008va}:
\be
V_{\mc{O}(\vo^{-3})}(\vo,\tau_3,\theta_3)  = \frac 83 \sqrt{\tau_3} A_3^2 a_3^2 \,\frac{e^{-2 a_3 \tau_3}}{\vo} + \cos(a_3 \theta_3)4 W_0 A_3 a_3 \tau_3 \,\frac{e^{-a_3\tau_3}}{\vo^2}+\frac{3 \xi W_0^2 }{4 g_s^{3/2}\vo^3}\,.
\label{LeadingV}
\ee
Notice that, at this order in the expansion, $V$ does not depend either on $\tau_1$ or on the axions $\theta_1$ and $\theta_2$. This is because the dominant contribution to the potential of $\tau_1$ arises only via string loops at order $\mc{O}(\vo^{-10/3})$ \cite{Cicoli:2008va}. On the other hand, $\theta_1$ and $\theta_2$ develop a potential at even more subleading order, $\mc{O}(\vo^{-p})$ with $p>10/3$, due to tiny non-perturbative effects \cite{Cicoli:2017zbx}.

The potential $V_{\mc{O}(\vo^{-3})}$ admits a supersymmetry-breaking AdS minimum at exponentially large volume (at first order in an expansion in $1/(a_3\tau_3)\ll 1$):
\be
 \langle \tau_3 \rangle=\,\left(
 \frac{\xi}{2} \right)^\frac23\frac{1}{g_s}\,,
\qquad \langle \vo\rangle  =\frac{3 W_0 \sqrt{\langle \tau_3\rangle }}{4 A_3 a_3 }\,e^{a_3 \langle \tau_3\rangle }\,,
\qquad \langle \theta_3\rangle = \frac{\pi}{a_3}\left(1+2\kappa_3\right)\quad\kappa_3\in\mathbb{Z}\,.
\label{mint3}
\ee

\subsubsection*{Moduli stabilisation at order $\mc{O}(\vo^{-10/3})$}
\label{potv103}

Without loss of generality and following \cite{Cicoli:2008gp}, we shall lift one of the three remaining flat directions by the inclusion of string loops while we shall ignore contributions from higher derivative corrections since, as studied in \cite{Broy:2015zba} and \cite{Cicoli:2016chb}, these $\mc{O}(\alpha'^3)$ $F^4$ terms do not modify qualitatively the inflationary dynamics but give rise just to some slight quantitative differences. The $g_s$ effects (\ref{Kgs}) generate a scalar potential of the form \cite{Cicoli:2007xp}:
\be
V_{g_s} = \left[\sum_i \left(g_s c_i^\KK\right)^2\partial^2_{\tau_i} K_{\rm tree}-2 K_{g_s}^\W\right]\frac{W_0^2}{\vo^2}\,.
\ee
In our case, focusing just on $\tau_1$ and $\tau_2$-dependent loop corrections, the resulting scalar potential becomes \cite{Cicoli:2008gp}:
\be
V_{\mc{O}(\vo^{-10/3})}(\tau_1)=\left(g_s^2\,\frac{A}{\tau_1^2}  -\frac{B}{\vo\sqrt{\tau_1}} +g_s^2\,\frac{C\tau_1}  {\vo^2}\right)\frac{W_0^2}{\vo^2},
\label{potwl}
\ee
where $A$, $B$, $C$ are given by:
\be
A=\left(c_1^\KK\right)^2>0,\qquad B=2 \alpha c^\W,\qquad C= 2\left(\alpha\,c_2^\KK\right)^2>0 \,.
\label{defC}
\ee
The parameters $c_1^\KK$, $c^\W$ and $c_2^\KK$ depend on the complex structure moduli which are fixed by background fluxes. Thus, from the string landscape point of view, these are tunable coefficients which can be adjusted due to phenomenological requirements. As shown in \cite{Burgess:2016owb}, $B<0$ leads to a prediction for $r$ ruled out by observations. Hence from now on we focus only on the case with $B>0$ which, when $g_s^4 \ll \left(\frac{c^\W}{4 c_1^\KK c_2^\KK}\right)^2$, yields a minimum for $\tau_1$ at:
\be
\langle \tau_1 \rangle \simeq g_s^{4/3}\lambda \langle\vo\rangle^{2/3}\qquad\text{with}\qquad 
\lambda \equiv \left(\frac{4A}{B} \right)^{2/3}\,, 
\label{tau1soln2}
\ee
justifying the $\vo^{-10/3}$ scaling of (\ref{potwl}). Notice that at the minimum the moduli scale as $\tau_3 \ll \tau_1 \ll \tau_2$ since:
\be
\langle\tau_3\rangle \ll g_s^{4/3}\,e^{\frac 23 a_3 \langle\tau_3\rangle} \simeq g_s^{4/3} \langle\vo\rangle^{2/3} \simeq 
\langle\tau_1\rangle \simeq g_s^2 \langle\tau_2\rangle \ll \langle\tau_2\rangle\,.
\label{Anis}
\ee

\subsubsection*{Moduli stabilisation at $\mc{O}(\vo^{-4/3} e^{-\vo^{2/3}})$}
\label{potv3p}

The leading contribution to the scalar potential from $T_1$ and $T_2$-dependent non-perturbative corrections to $W$ is given by (we neglect mixed terms since they have a double exponential suppression):
\be
V_{\mc{O}(\vo^{-4/3} e^{-\vo^{2/3}})}(\theta_1,\theta_2)=\frac{4 W_0}{\vo^2}\left[A_1 a_1 \tau_1 e^{-a_1 \tau_1}  \cos(a_1 \theta_1)
+ A_2 a_2 \tau_2 e^{-a_2 \tau_2} \cos(a_2 \theta_2)\right].
\label{Vp}
\ee
This potential turns out to be more suppressed than the ones analysed before since it scales as $\tau_i e^{-a_i \tau_i} \vo^{-2} \sim \vo^{-4/3} e^{-\vo^{2/3}}$ but it is still the dominant potential for $\theta_1$ and $\theta_2$ which are stabilised at:
\be
\langle \theta_1 \rangle = \frac{\pi}{a_1}\left(1+2\kappa_1\right)\qquad\text{and}\qquad\langle \theta_2\rangle=\frac{\pi}{a_2}\left(1+2\kappa_2\right)\quad\kappa_1\,,\kappa_2\in\mathbb{Z}\,.
\ee 
We stress again that this minimum is AdS and so an additional term is needed to uplift the potential to a nearly Minkowski vacuum. As recently reviewed in \cite{Cicoli:2018kdo}, there are several mechanisms to achieve a dS vacuum via either anti-branes \cite{Kachru:2003aw}, T-branes \cite{Cicoli:2015ylx}, non-perturbative effects at singularities \cite{Cicoli:2012fh} or non-zero F-terms of the complex structure moduli \cite{Gallego:2017dvd}.

\subsubsection*{Mass spectrum}

At leading $\vo^{-3}$ order, the scalar potential depends just on three moduli: the blow-up mode $\tau_3$ and its associated axion $\theta_3$, and a particular combination of $\tau_1$ and $\tau_2$ corresponding to the overall volume $\vo$. Hence, at this level of approximation, only these three fields acquire masses of order:
\be
m_{\tau_3}^2 \sim m_{\theta_3}^2 \sim m_{3/2}^2\qquad\text{and}\qquad m_\vo^2\sim \frac{m_{3/2}^2}{\vo}\,, 
\label{masses}
\ee
where the gravitino mass scales as $m_{3/2}^2 = e^K\,|W|^2\sim \frac{W_0^2}{\vo^2}\,M_p^2$. As shown in \cite{Cicoli:2010ha,Burgess:2010bz}, the kinetic terms for the K\"ahler moduli can be diagonalised at leading order in $1/\vo$ giving:
\be
\tau_1 = e^{\sqrt{2}k \,\chi+2 k\,\phi}
\qquad\text{and}\qquad
\tau_2 = e^{\sqrt{2}k \,\chi-k \,\phi}\qquad\text{with}\qquad k\equiv \frac{1}{\sqrt{3}}\,,
\label{volTrans}
\ee
which implies: 
\be
\chi= \sqrt{2}k\ln\vo \qquad\text{and}\qquad
\phi=\frac{3k}{2} \ln \tau_1 - k \ln\vo\,.
\label{vol2}
\ee
Hence $\chi$ corresponds to $\vo$, and so it is the direction fixed at leading order, while $\phi$ corresponds to the combination of $\tau_1$ and $\tau_2$ which remains flat. This remaining flat direction is lifted at subleading order by $g_s$ perturbative effects and plays the r\^ole of the inflaton. The mass of $\phi$ at the minimum then sets the value of the Hubble constant during inflation $H_{\rm inf}$:
\be
\left.m_\phi^2\right|_{\rm min} \sim H_{\rm inf}^2 \sim \frac{m_{3/2}^2}{\vo^{4/3}}\,.
\label{Hubble}
\ee
However during inflation the inflaton, which from (\ref{vol2}) we can parametrise as $\tau_1$, is displaced from its minimum. Thus its potential (\ref{potwl}) becomes exponentially suppressed in terms of the canonically normalised inflaton $\phi$ which turns out to be very light since:
\be
\left.m_\phi^2\right|_{\rm inf} \sim \left.m_\phi^2\right|_{\rm min}\,e^{-k\hat\phi}\ll H_{\rm inf}^2 \,,
\ee
where $\hat\phi$ denotes the shift of $\phi$ from its minimum. Hence $\phi$ drives a period of slow-roll inflation while the other moduli sit approximately at their minima (since they are heavy). For $\vo\gg 1$, the shifts of the minima of the heavy fields are negligible and the dynamics is effectively single-field. The pictorial view of the inflationary process is rather simple. Inflation starts in a region where the fibre modulus $\tau_1$ is much larger than the base modulus $\tau_2$, while during the inflationary evolution the fibre decreases while the base becomes larger keeping $\vo$ approximately constant. At the end of inflation the inflaton reaches its minimum where, according to (\ref{Anis}), the Calabi-Yau is anisotropic since the base is much larger than the fibre.

On the other hand the two axions $\theta_1$ and $\theta_2$ are almost massless since they are stabilised at order $\mc{O}(\vo^{-4/3}\,e^{-\vo^{2/3}})$:
\be
m_{\theta_i}^2 \sim \tau_i^2 \, V_{\theta_i \theta_i} \sim m_{3/2}^2 \tau_i^3 e^{-a_i \tau_i} \sim e^{-\vo^{2/3}} M_p^2 \sim 0 \qquad \forall i=1,2\,.
\ee
Notice moreover that, since we will realise a chiral visible sector via a stack of D7-branes wrapped around the fibre divisor $D_1$, the prefactors $A_1$ and $A_2$ of the non-perturbative contributions in (\ref{W0pot}) could be vanishing due to the well-known tension between moduli stabilisation and chirality \cite{Blumenhagen:2007sm}. In this case $\theta_1$ and $\theta_2$ would even be lighter. Given that these pseudoscalar fields are in practice massless, their contribution to the total dark matter density is negligible, and so present isocurvature fluctuation bounds \cite{Akrami:2018odb} can be easily satisfied.

\subsection{Single-field inflation}
\label{SFI}

\begin{figure}[!htbp]
\centering
\includegraphics[scale = 0.8]{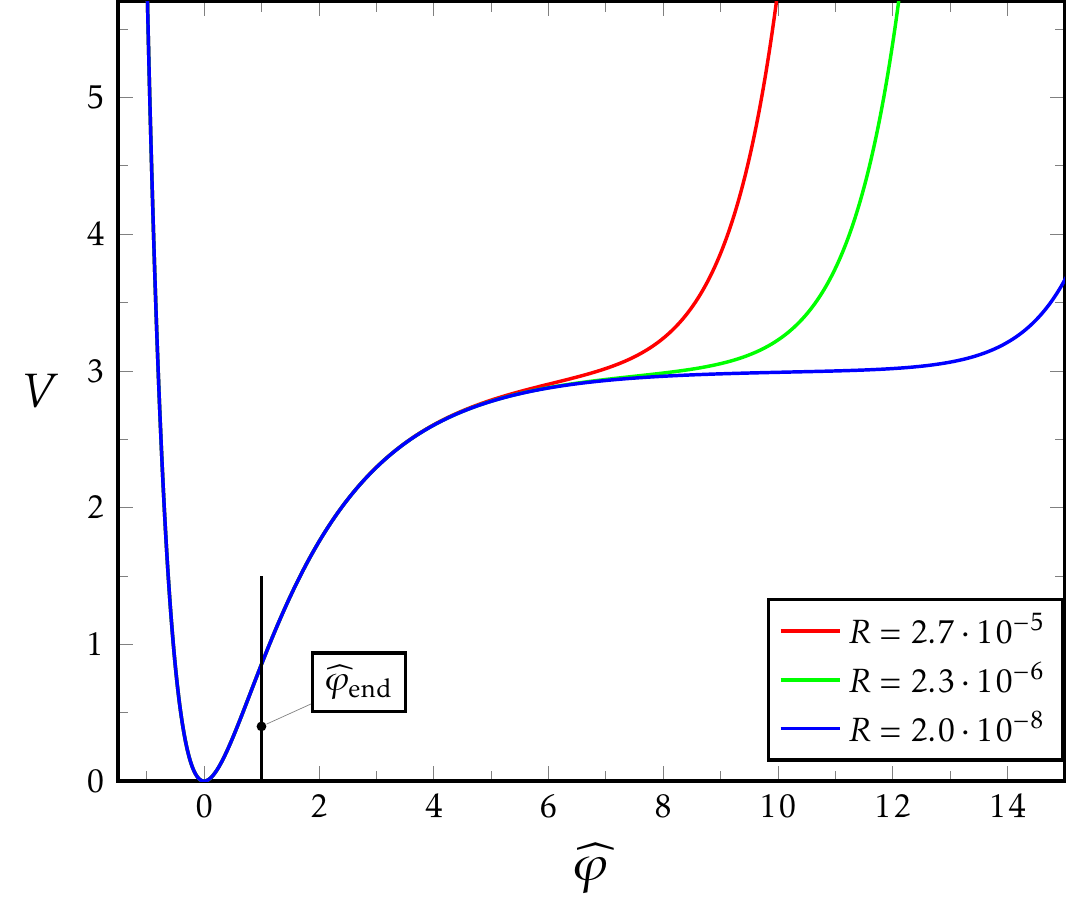}
\caption{Inflationary potential for different values of $R$ setting $V_0 = 1$. The plot shows also the end point of inflation used in the following analysis.}
\label{Fig1}
\end{figure}

The inflationary dynamics can be studied by expanding $\tau_1$ in terms of its canonically normalised counterpart using (\ref{volTrans}), and then writing $\phi = \langle \phi \rangle + \hat \phi $ to denote the shift from the minimum:
\be
 \tau_1= \langle\tau_1\rangle \,e^{2k\,\hat\phi} \simeq \lambda g_s^{4/3} \langle\vo\rangle ^{2/3}\,e^{2k\,\hat\phi}\,.
\ee
After expressing $\tau_1$ in terms of the canonically normalised field $\hat \phi$, the inflationary potential (\ref{potwl}) looks like:\footnote{We set $M_p=1$ and introduce the correct overall normalisation factor $g_s/8\pi$ obtained from dimensional reduction~\cite{Burgess:2010bz}.}
\be
 V_{\rm inf}(\hat\phi) =V_0 \,\left[3-4 \,e^{-k\, \hat \phi }
 +  e^{-4k\,  \hat \phi}+ R\left(e^{2k \,\hat \phi} - 1\right)\right] \,,
\label{Inflationpot}
\ee
where:
\be
V_0 = \frac{g_s^{1/3}\,W_0^2 \,A}{8\pi\,\lambda^2 \,\langle\vo\rangle^{10/3}} \qquad \text{and}\qquad 
R \equiv 16 g_s^4 \frac{A C}{B^2} = 2g_s^4  \left(\frac{c_1^\KK c_2^\KK}{c^\W}\right)^2 \ll 1\,.
\label{V0}
\ee
In the previous expression we added $\hat\phi$-independent contributions to the scalar potential needed to uplift the original AdS vacuum to a nearly Minkowski one which can come from several sources of positive energy like anti-branes \cite{Kachru:2003aw}, T-branes \cite{Cicoli:2015ylx}, non-perturbative effects at singularities \cite{Cicoli:2012fh} or non-vanishing F-terms of the complex structure moduli \cite{Gallego:2017dvd}. Moreover notice that the prefactor of (\ref{Inflationpot}) sets both the inflaton mass at the minimum and the Hubble scale during inflation $H_{\rm inf}^2\simeq V_{\rm inf}/3$: 
\be
\left.m_\phi^2\right|_{\rm min} = \left.V_{\rm inf}''\right|_{\hat\phi=0} = 4 V_0 \left(1+\frac{R}{3}\right) 
\simeq \frac{m_{3/2}^2}{\vo^{4/3}} \simeq H_{\rm inf}^2\,,
\label{mphimin}
\ee
in perfect agreement with (\ref{Hubble}). Clearly, the inflaton mass quickly becomes exponentially smaller than $H_{\rm inf}$ for $\hat\phi>0$. Fig.~\ref{Fig1} shows the inflationary potential for different values of $R$. 

\subsubsection*{Inflationary predictions}

Starting from the inflationary potential (\ref{Inflationpot}), the slow-roll parameters become:
\bea
 && \epsilon(\hat\phi,R) \equiv \frac12 \left(\frac{V_{\rm inf}'}{V_{\rm inf}}\right)^2 =
 \frac23\frac{\left(2 \,e^{-k\, \hat \phi } -2  \,e^{-4k\,  \hat \phi}+R \,e^{2k \,\hat \phi} \right)^2}
{\left(3-R+  e^{-4k\,  \hat \phi}-4 \,e^{-k\, \hat \phi } +R\,e^{2k \,\hat \phi } \right)^2} \,,  
\label{eps} \\
 && \eta(\hat\phi,R) \equiv \frac{V_{\rm inf}''}{V_{\rm inf}} =
 \frac43\frac{4  \,e^{-4k\,  \hat \phi}-  \,e^{-k\, \hat \phi } +R \,e^{2k \,\hat \phi}}
{\left(3-R +  e^{-4k\,  \hat \phi}-4 \,e^{-k\, \hat \phi } +R\,e^{2k\, \hat \phi} \right)} \,.
\label{eta}
\eea
The slow-roll parameter $\eta$ vanishes at the two inflection points $\hat\phi_{\rm ip}^{(1)}\simeq k\ln 4\simeq 0.8$ where the two negative exponentials compete with each other, and $\hat\phi_{\rm ip}^{(2)}\simeq - k\ln R\gg \hat\phi_{\rm ip}^{(1)}$ for $R\ll 1$ where the positive exponential becomes comparable in size with $e^{-k\hat\phi}$. The slow-roll parameter $\epsilon$ at $\hat\phi_{\rm ip}^{(1)}$ becomes $\epsilon_{\rm ip}^{(1)}\simeq 3/2$, signaling that inflation ends close to the first inflection point. In fact, $\epsilon\simeq 1$ around $\hat\phi_{\rm end}\simeq 1$, independently of the microscopic parameters since the term proportional to $R$ can be neglected in the vicinity of the minimum. As in \cite{Cicoli:2008gp}, there is an inflationary plateau to the right of the first inflection point and inflation takes place for field values in the window $\hat\phi_{\rm ip}^{(1)}<\hat\phi_{\rm end}\lesssim\hat\phi\lesssim \hat\phi_{\rm ip}^{(2)}$ since the spectral index is always too blue for $\hat\phi\gg \phi_{\rm ip}^{(2)}$.

The number of efoldings between the point of horizon exit $\hat\phi_*$ and the end of inflation is then computed as:
\be
N_e (\hat\phi_*,R)= \int_{\hat\phi_{\rm end}}^{\hat\phi_*} U(\hat\phi,R)\,d\hat\phi\,,
\label{Nef}
\ee
with:
\be
U(\hat\phi,R)= \frac{1}{\sqrt{2\epsilon(\hat\phi,R)}} \simeq 
\frac{\sqrt{3}}{2}\frac{\left(3-4 \,e^{-k\, \hat \phi } 
+R\,e^{2k \,\hat \phi } \right)}{\left(2 \,e^{-k\, \hat \phi } +R \,e^{2k \,\hat \phi} \right)}\,,
\label{Uphi}
\ee
where we have neglected the term in $V_{\rm inf}$ proportional to $e^{-4k\,  \hat \phi}$ since slow-roll inflation occurs in the region with $\hat\phi>\hat\phi_{\rm ip}^{(1)}$. 

The main cosmological observables we are interested in are the scalar spectral index $n_s$ and the tensor-to-scalar ratio $r$ which have to be evaluated at horizon exit as:
\be
n_s (\hat\phi_*) = 1 +2\eta (\hat\phi_*)  -6\epsilon(\hat\phi_*) \qquad\text{and}\qquad r(\hat\phi_*) =16\epsilon (\hat\phi_*) \,.
\label{nsr}
\ee
We would like now to solve the integral (\ref{Nef}), and then invert the relation $N_e=N_e(\hat\phi_*,R)$ to obtain $\phi_*=\phi_*(N_e,R)$. Substituting this relation into $n_s (\hat\phi_*)$ and $r (\hat\phi_*)$, we would end up with $n_s =n_s(N_e,R)$ and $r=r(N_e,R)$, where $N_e$ depends on the post-inflationary evolution and $R$ is a tunable parameter which can be adjusted to reproduce the correct Planck value of the spectral index once the prediction for $\Delta N_{\rm eff}$ is determined. 

However the integral (\ref{Nef}) cannot be solved analytically. We shall therefore consider two simplified cases where an approximated analytical solution can be provided:
\ben
\item $\hat\phi_*\ll \hat\phi_{\rm ip}^{(2)}$: In this case we can set $R=0$ and (\ref{Uphi}) simplifies to:
\be
U(\hat\phi) \simeq 
\frac{\sqrt{3}}{4}\left(3\,e^{k \hat \phi }-4 \right)\,.
\label{Uphisimpl}
\ee
and so the integral in (\ref{Nef}) can be solved exactly yielding:
\be
N_e (\hat\phi_*) \simeq \frac94 e^k \left(e^{k \Delta\hat\phi} - 1\right)  -3k \Delta\hat\phi\,,
\ee
where $\Delta\hat\phi=\hat\phi_*-\hat\phi_{\rm end}\simeq\hat\phi_*-1$. This equation can be inverted iteratively to give:
\be
\phi_* (N_e) = \sqrt{3}\ln\left(f(N_e)+\frac43\ln f(N_e)\right)
\quad\text{with}\quad f(N_e) \equiv \frac49 N_e+e^k-\frac{4k}{3}\,.
\label{phistNe}
\ee
In this region where $\hat\phi_{\rm ip}^{(1)}< \hat\phi_*\ll \hat\phi_{\rm ip}^{(2)}$, the slow-roll parameters (\ref{eps}) and (\ref{eta}) can be well approximated as:
\be
\epsilon(\hat\phi) \simeq \frac32 \,\eta(\hat\phi)^2 \qquad\text{with}\qquad \eta(\hat\phi) \simeq
- \frac43\frac{\,e^{-k\, \hat \phi }}
{\left(3-4 \,e^{-k\, \hat \phi }\right)}\,,
\label{etaapprox}
\ee
which implies an interesting relation between the cosmological observables $n_s$ and $r$: 
\be
n_s -1= 2\eta  -9 \,\eta^2 \simeq 2\eta \qquad\text{and}\qquad r =24 \,\eta^2 \quad\Rightarrow\quad  r\simeq 6\left(n_s  -1\right)^2\,.
\label{nsr}
\ee
If we substitute (\ref{etaapprox}) into (\ref{nsr}) with $\phi_*$ given by (\ref{phistNe}), the spectral index can be expressed as a function of the number of efoldings as:
\be
n_s(N_e) = 1 +\frac{8}{4 \left(3 + \sqrt{3} - N_e\right)-9\, e^k - 12 \ln \left(e^k + \frac49 N_e\right)}\,,
\ee
while the tensor-to-scalar ratio is given from (\ref{nsr}) by $r(N_e)\simeq 6\left(n_s(N_e)  -1\right)^2$. 

\begin{figure}[!htbp]
\centering
\includegraphics[scale = 0.6]{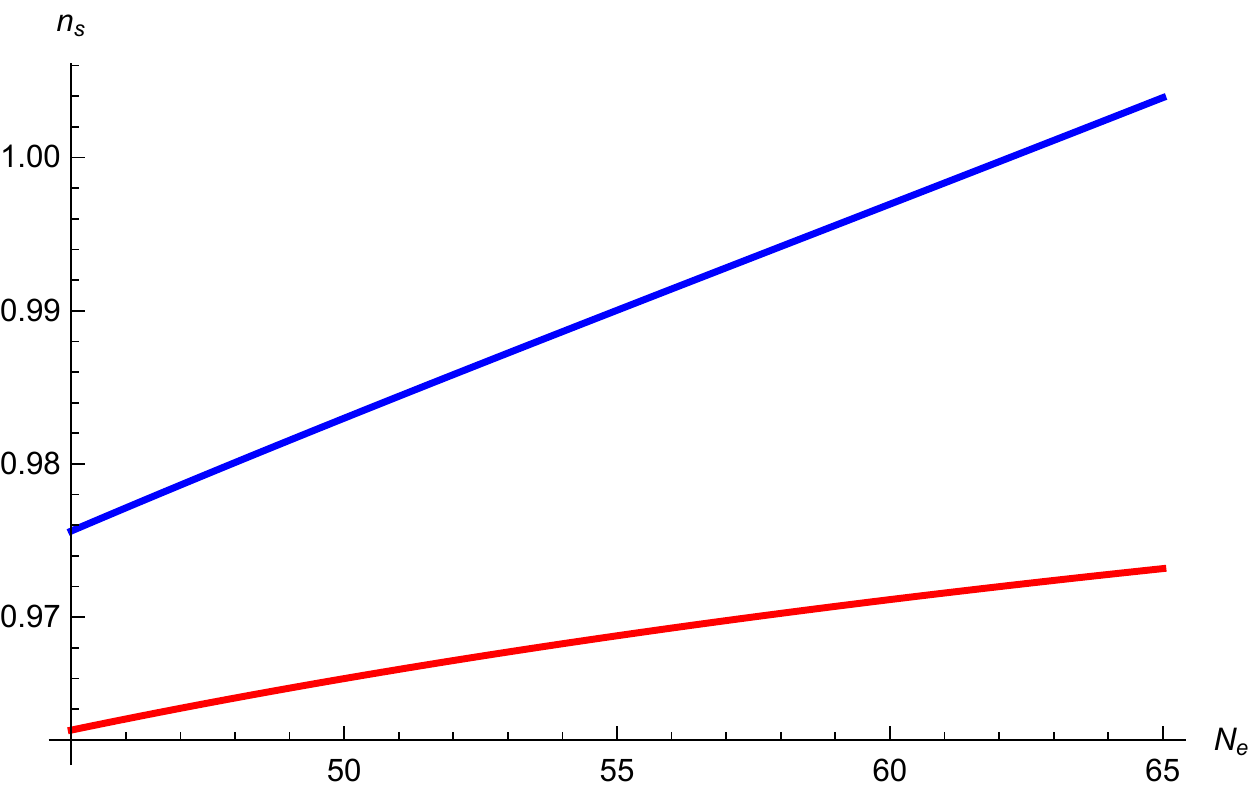}
\includegraphics[scale = 0.6]{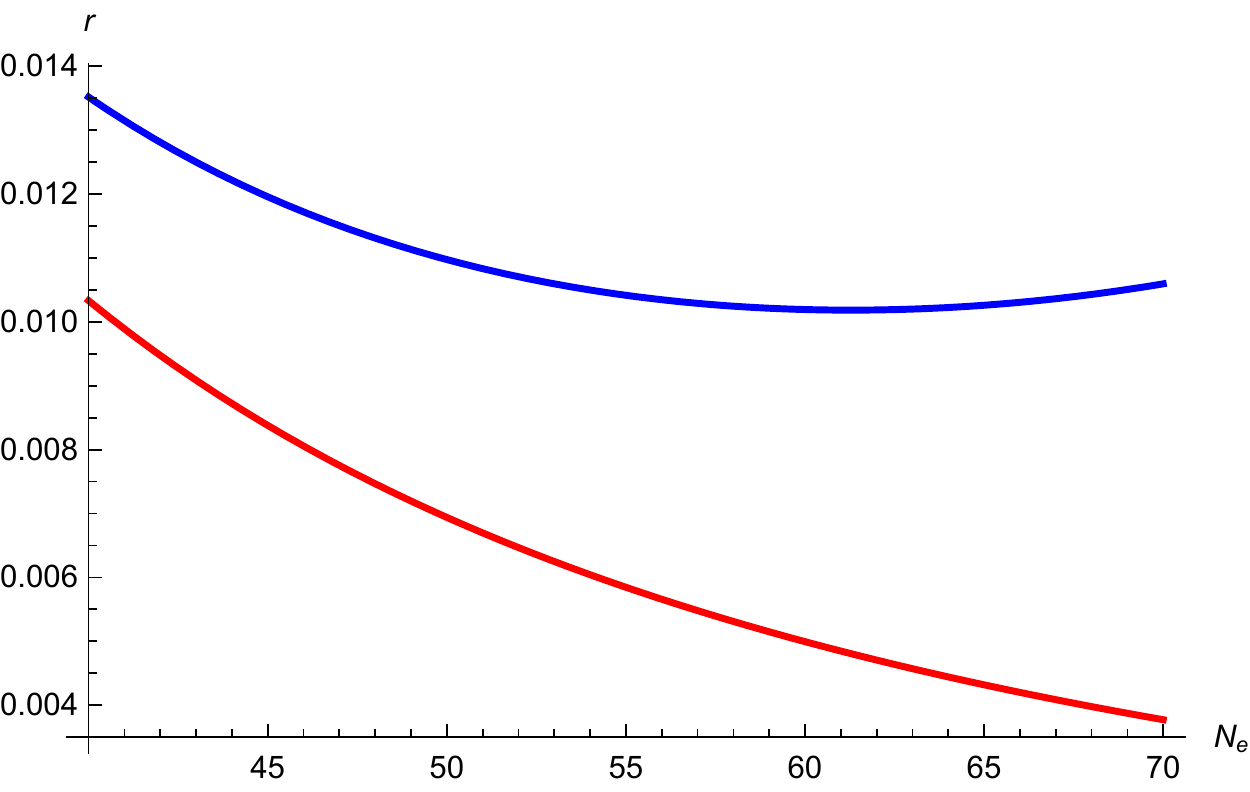}
\caption{Scalar spectral index and tensor-to-scalar ratio as a function of the number of e-foldings for the case with $R\ll 1$ (red lines) and the case where the positive exponential in $V_{\rm inf}$ cannot be neglected (blue lines fixing $R=2.7\cdot 10^{-5}$).}
\label{Fig1b}
\end{figure}

\begin{figure}[!htbp]
\centering
\includegraphics[scale = 0.6]{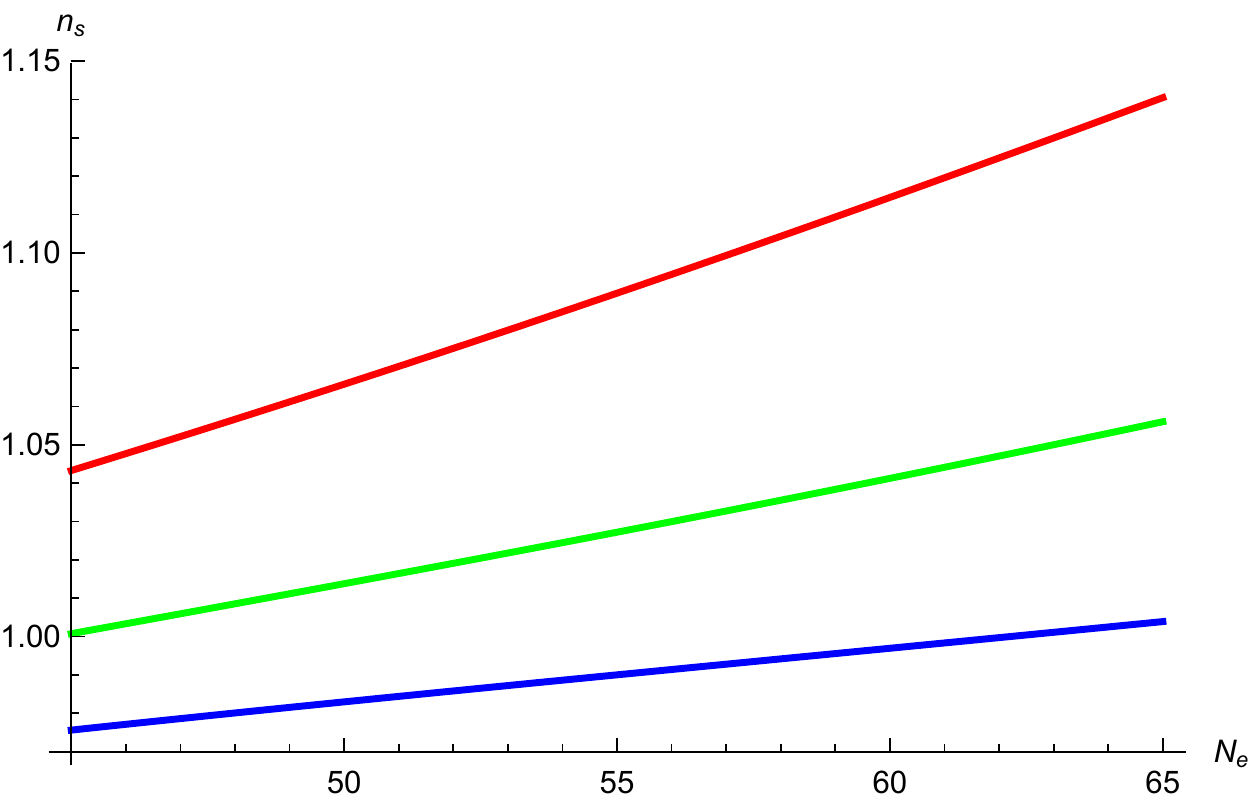}
\includegraphics[scale = 0.6]{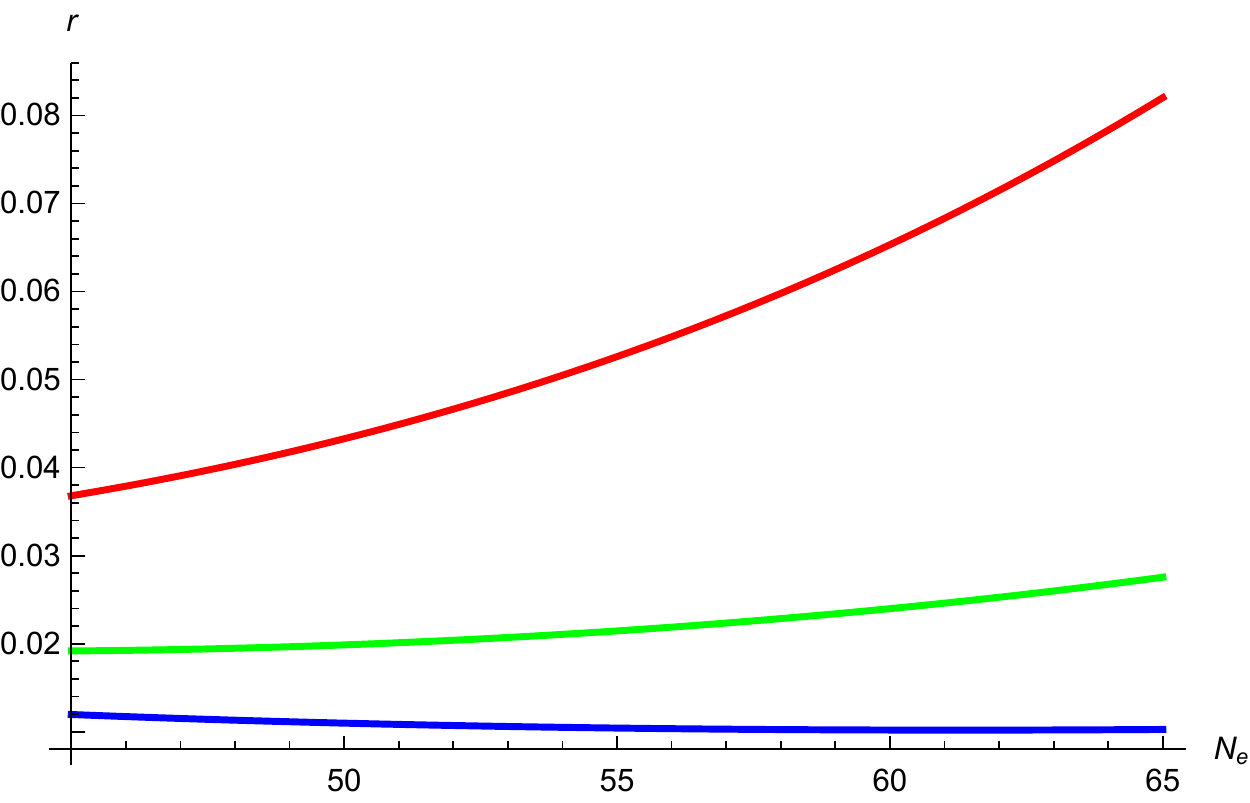}
\caption{Scalar spectral index and tensor-to-scalar ratio as a function of the number of e-foldings for different values of $R$ in the case where the positive exponential in $V_{\rm inf}$ cannot be neglected (blue lines for $R=R_0 \equiv 2.7\cdot 10^{-5}$, green lines for $R=2.5 R_0$ and red lines for $R=5 R_0$).}
\label{Fig2b}
\end{figure}

\item $\hat\phi_*\lesssim \hat\phi_{\rm ip}^{(2)}$: In this case we need to keep the term proportional to $R$ but the integrand (\ref{Uphi}) can be expanded in power series as:
\be
U(\hat\phi,R)\simeq \frac{\sqrt{3}}{2}\left[1 +\frac32 \left(e^{k\hat\phi}-2\right) \sum_{n= 0}^\infty (-1)^n e^{n\hat\phi/k} \left(\frac{R}{2}\right)^n\right] \,,
\label{UphiSimpl}
\ee
and so the integral in (\ref{Nef}) now admits the following solution:
\bea
N_e (\hat\phi_*,R) &\simeq& \frac94 e^k \left(e^{k \Delta\hat\phi} - 1\right)  -3k \Delta\hat\phi \nn \\
 &+& \frac34 \sum_{n=1}^\infty (-1)^n \left[\frac{e^{k+\frac{n}{k}}}{k^2+n} \left(e^{\left(k+\frac{n}{k}\right)\Delta\hat\phi}-1\right)
-\frac{2\,e^{\frac{n}{k}}}{n}  \left(e^{\frac{n}{k}\Delta\hat\phi}-1 \right)\right]   \left(\frac{R}{2}\right)^n \,. \nn
\eea
We can again invert this equation iteratively, obtaining:
\bea
\hat\phi_* (N_e,R)   &\simeq& \sqrt{3}\ln\left\{f(N_e)+\frac43\ln f(N_e)\right.  \label{phistNeR} \\
 &- & \left.\frac13 \sum_{n=1}^\infty (-1)^n 
\left[\frac{1}{k^2+n} \left(f(N_e)^{1+\frac{n}{k^2}}-e^{k+\frac{n}{k}}\right)
-\frac{2}{n}  \left( f(N_e)^{\frac{n}{k^2}}-e^{\frac{n}{k}} \right)\right]   \left(\frac{R}{2}\right)^n\right\}. \nn
\eea
In this region where $\hat\phi_{\rm ip}^{(1)}\ll \hat\phi_*\lesssim \hat\phi_{\rm ip}^{(2)}$, the slow-roll parameters (\ref{eps}) and (\ref{eta}) can be well approximated as:
\be
\epsilon(\hat\phi,R) \simeq \frac32 \eta^2\left(1+\frac{R}{2} e^{\hat\phi/k}\right)^2\sum_{n=0}^\infty (n+1) R^n e^{n\hat\phi/k}\,,
\ee
where:
\be
\eta(\hat\phi,R) \simeq  -\frac49\left(e^{-k\, \hat \phi } -R \,e^{2k \,\hat \phi}\right).
\label{etaapprox2}
\ee
This gives a modified relation between the cosmological observables $n_s$ and $r$ since: 
\be
n_s -1= 2\eta  - 9 \eta^2\left(1+\frac{R}{2} e^{\hat\phi/k}\right)^2\sum_{n=0}^\infty (n+1) R^n e^{n\hat\phi/k}\simeq 2\eta\,,
\label{nsrnew}
\ee
and:
\be
r \simeq 6\left(n_s  -1\right)^2\left(1+\frac{R}{2} e^{\hat\phi/k}\right)^2\sum_{n=0}^\infty (n+1) R^n e^{n\hat\phi/k} \quad\Rightarrow\quad  r> 6\left(n_s  -1\right)^2\,,
\label{nsr2}
\ee
implying that, for fixed $n_s$, this case gives a larger $r$ with respect to the first case. This result is not surprising since in the case where the positive exponential contribution to $V_{\rm inf}$ is not negligible, the inflationary potential turns out to be steeper. If we substitute (\ref{etaapprox2}) into (\ref{nsrnew}) and (\ref{nsr2}) with $\phi_*$ given by (\ref{phistNeR}), we can obtain the functions $n_s=n_s(N_e,R)$ and $r=r(N_e,R)$. 
\een
Fig.~\ref{Fig1b} shows the spectral index and the tensor-to-scalar ratio as a function of $N_e$ for the two cases discussed above. The red lines represent case $1$ where the positive exponential in $V_{\rm inf}$ is negligible throughout the whole inflationary dynamics (setting $R=0$) while the blue lines correspond to case $2$ (setting $R=2.7\cdot 10^{-5}$). Notice that, for the same number of e-foldings, case $1$ gives smaller $n_s$ and $r$. In particular, in case $1$, $n_s$ cannot be larger than about $0.97$ for $N_e\leq 65$. On the other hand, Fig.~\ref{Fig2b} focuses on case $2$ and shows $n_s=n_s(N_e,R)$ and $r=r(N_e,R)$ for different values of $R$. Notice that larger value of $R$ give larger $n_s$ and $r$. 

\begin{figure}[!htbp]
\centering
\includegraphics[scale = 0.7]{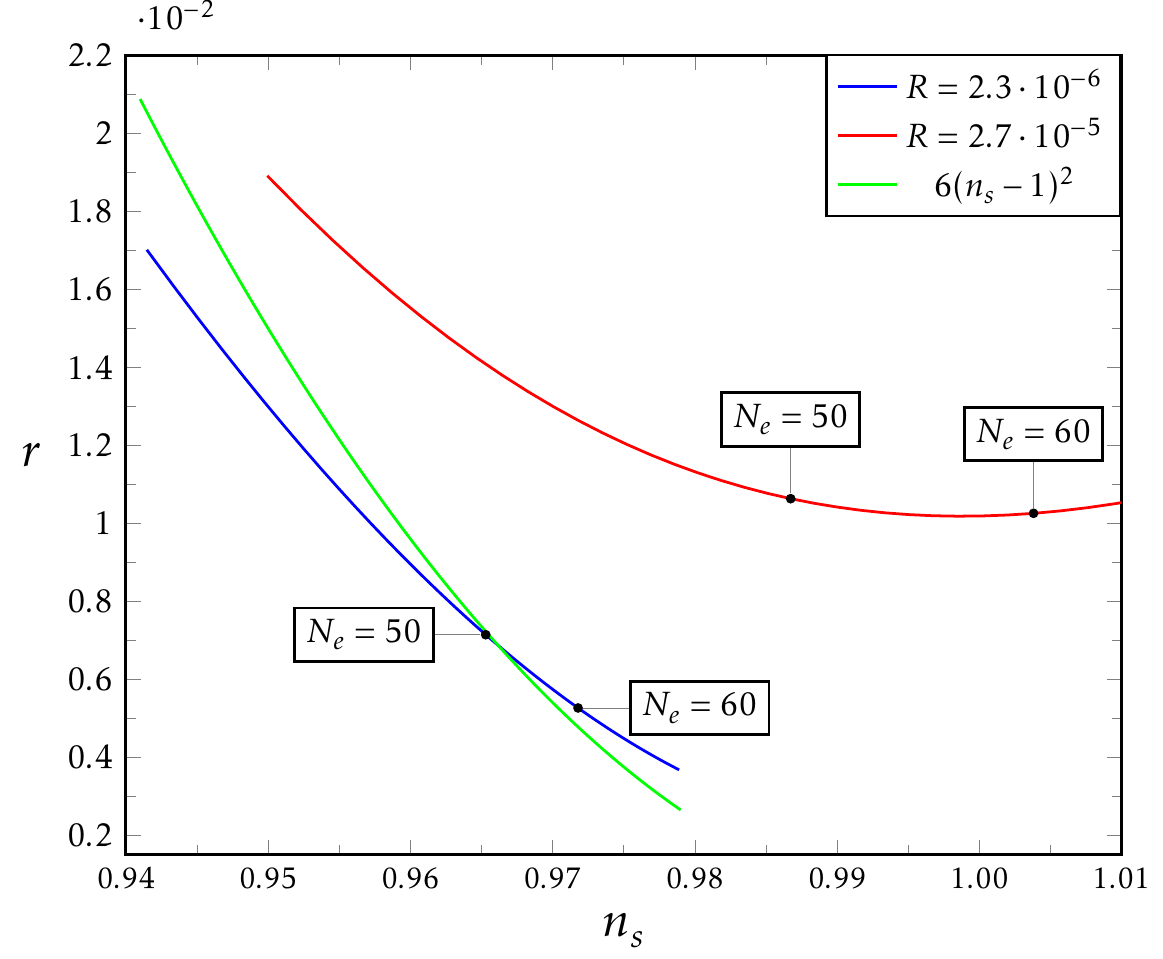}
\caption{Tensor-to-scalar ratio as a function of the spectral index for two different values of $R$ (red and blue lines). The green curve represents the relation $r = 6(n_s -1)^2$.}
\label{Fig3}
\end{figure}

Fig.~\ref{Fig3} shows instead $r$ versus $n_s$ for two different values of the parameter $R$. The green curve represents the relation $r=6(n_s-1)^2$ which we showed to be a good approximation for case $1$ with $R=0$. Interestingly, values of $R$ of order $R=2.3\cdot 10^{-6}$ fall under case $1$ while $R=2.7\cdot 10^{-5}$ already belongs to case $2$ where the relation $r=6(n_s-1)^2$ is violated.

As we have seen, a crucial quantity to make definite cosmological predictions is the number of e-foldings $N_e$ which depends on the post-inflationary evolution of our Universe and the inflationary energy scale. If the Universe is radiation dominated all the way from reheating to matter-radiation equality, and the reheating epoch is characterised by a generic equation of state $p = w_{\rm rh}\rho$ and a reheating temperature $T_{\rm rh}$, $N_e$ is given by \cite{Planck:2013jfk}:
\be
N_e \simeq 57 + \frac14 \ln r + \frac 14 \ln\left(\frac{\rho_*}{\rho_{\rm end}}\right)+\frac{1-3w_{\rm rh}}{12(1+3w_{\rm rh})}
\ln\left(\frac{\pi^2}{45}\,g_*(T_{\rm rh})\right)-\frac{1-3w_{\rm rh}}{3(1+3w_{\rm rh})}\ln\left(\frac{M_{\rm inf}}{T_{\rm rh}}\right), \nn
\ee
where $\rho_{\rm end} \simeq M_{\rm inf}^4=  V_{\rm end}$. In our case, the inflaton oscillates around its minimum after the end of inflation and its decay drives reheating leading to the production of SM particles. Hence the reheating epoch is dominated by non-relativistic matter with $w_{\rm rh}=0$. Moreover, the Hubble parameter is approximately constant during inflation, implying that $\rho_*\simeq \rho_{\rm end}$. Therefore the previous expression simplifies to (the term proportional to $g_*(T_{\rm rh})= 106.75$, for $T_{\rm rh}$ larger than the top mass, gives rise to a negligible contribution): 
\be
N_e \simeq 57 + \frac14 \ln r -\frac13\ln\left(\frac{M_{\rm inf}}{T_{\rm rh}}\right).
\label{Ne0}
\ee
As can be seen from Fig. \ref{Fig1b}, \ref{Fig2b} and \ref{Fig3}, $0.005\lesssim r\lesssim 0.02$ for $0.95\lesssim n_s\lesssim 1.00$, which gives $\frac14 \ln r\simeq -1$. Thus the number of efoldings (\ref{Ne0}) can finally be rewritten in terms of $M_{\rm inf}$ and $T_{\rm rh}$ as: 
\be
N_e \simeq 56 -\frac13\ln\left(\frac{M_{\rm inf}}{T_{\rm rh}}\right).
\label{Ne1}
\ee
Using the measured amplitude of the density perturbations, $A_s \simeq 2.3\cdot 10^{-9}$ \cite{Planck:2013jfk}, and under the approximation $V_*\simeq V_{\rm end}$, the inflationary scale $M_{\rm inf}$ can be written in terms of the tensor-to-scalar ratio as:
\be
M_{\rm inf} \simeq V_*^{1/4} = \left(\frac{3\pi^2}{2}\,A_s\,r\right)^{1/4} M_p \simeq \left(\frac{r}{0.1}\right)^{1/4} M_\GUT\,,
\label{Minf}
\ee
with $M_\GUT \equiv 1.85\cdot 10^{16}$ GeV. Eq. (\ref{Minf}) shows that $M_{\rm inf}\simeq 10^{16}$ GeV for $0.005\lesssim r\lesssim 0.02$. Moreover it sets a constraint on the overall scale $V_0$ of the inflationary potential (\ref{Inflationpot}) in order to match the observed amplitude of the density perturbations at horizon exit since $V_0 \simeq M_{\rm inf}^4$. Given the expression (\ref{V0}) for $V_0$, this matching can be guaranteed by a proper choice of underlying parameters. In particular, for natural values $A=1$ and $\lambda=2$ (which can come from $c_1^\KK = c^\W=k_{122}=1$), $g_s=0.1$ and $W_0=15$, (\ref{Minf}) implies $\langle \vo \rangle \simeq 10^3$ for $0.005\lesssim r\lesssim 0.02$.

Let us point out that (\ref{Minf}) fixes also the Hubble scale during inflation and the mass of the inflaton at the minimum since from (\ref{mphimin}) we have:
\be
H_{\rm inf}\simeq \left.m_\phi\right|_{\rm min} \simeq \frac{M_{\rm inf}^2}{M_p} \simeq  5\cdot 10^{13}\,\text{GeV}\,.
\label{massphi}
\ee
Furthermore, from (\ref{mphimin}) we see that the gravitino mass turns out to be of order:
\be
m_{3/2}\simeq H_{\rm inf} \vo^{2/3} \simeq M_{\rm inf} \left(\frac{M_{\rm inf}}{M_p}\right)^{1/5}\simeq 5\cdot 10^{15}\,\text{GeV}\,.
\ee
In string models where supersymmetry is broken in the bulk by non-zero F-terms of the K\"ahler moduli, the soft terms generated by gravity mediation are of order the gravitino mass, $M_{\rm soft}\simeq m_{3/2}$, implying that in our inflationary scenario, squarks, sleptons and gauginos acquire masses of order the GUT scale. Interestingly, in the simplest version of the MSSM with universal soft terms, such a high value of $M_{\rm soft}$ is incompatible with a $125$ GeV Higgs \cite{Bagnaschi:2014rsa}. However, a correct Higgs mass can easily be obtained by considering more general supersymmetric models with non-universal soft terms \cite{Ellis:2017erg} or in simple extensions of the MSSM like the NMSSM \cite{Zarate:2016jch}. Another possibility would be to consider a sequestered visible sector on D3-branes at singularities \cite{Conlon:2008wa,Cicoli:2012vw,Blumenhagen:2009gk, Aparicio:2014wxa} but, as shown in \cite{Angus:2014bia}, this case is not viable since it would lead to dark radiation overproduction. Finally one could also consider non-standard mechanisms to generate the density perturbations which could in principle give rise to larger Calabi-Yau volumes. However existing non-standard LVS inflationary scenarios, like the curvaton model presented in \cite{Burgess:2010bz} or the modulated reheating setup described in \cite{Cicoli:2012cy}, still feature overall volumes of the same order of magnitude.

\section{Reheating after fibre inflation}
\label{SecReheat}

In this section, after computing the inflaton coupling to both hidden and visible sector degrees of freedom, we shall study its perturbative decay after the end of inflation, obtaining the reheating temperature $T_{\rm rh}$ and the number of extra neutrino-like species $\Delta N_{\rm eff}$. Using (\ref{Ne1}) $T_{\rm rh}$ fixes $N_e$, while $\Delta N_{\rm eff}$, when used as a prior for Planck data, fixes the value of $n_s$ which has to be reproduced. Subsequently the function $n_s=n_s(N_e,R)$ will allow us to fix the microscopic flux-dependent parameter $R$, while $r=r(N_e,R)$ will give a definite prediction for the tensor-to-scalar ratio $r$. 

\subsection{Inflaton interactions}

\subsubsection{Hidden sector fields}

Hidden sector degrees of freedom are a generic feature of string compactifications since consistency conditions like tadpole cancellation can force the presence of hidden stacks of branes with light open string modes. However the presence of these light modes is model-dependent since in some constructions the cancellation of all anomalies might be achieved without the need to add hidden sector branes (see for example the explicit Calabi-Yau realisations of Fibre Inflation models presented in \cite{Cicoli:2016xae,Cicoli:2017axo}), or the hidden sector might be a pure super Yang-Mills theory that develops a mass gap, and so does not feature any light particle \cite{Cicoli:2010ha,Cicoli:2010yj}. Moreover, the hidden sector dynamics could be characterised by a Higgs-like mechanism which renders all hidden fields heavier than the inflaton. 

On the other hand, ultra-light closed string axions are a generic source of hidden sector fields \cite{Svrcek:2006yi, Conlon:2006tq, Arvanitaki:2009fg, Cicoli:2012sz}. In our model, the two ultra-light bulk axions $\theta_1$ and $\theta_2$ can potentially be produced from the inflaton decay after the end of inflation and can then give rise to positive contributions to the number of extra neutrino-like species $\Delta N_{\rm eff}$. In what follows we shall focus just on these model-independent hidden sector degrees of freedom and compute the inflaton coupling to these ultra-light modes.

These couplings arise from the kinetic Lagrangian which looks like:
\be
\mc{L}_{\rm kin} = \frac{\partial^2 K}{\partial T_i \partial \bar{T}_j}\, \partial^\mu T_i \partial_\mu \bar{T}_j = 
\frac14 \frac{\partial^2 K}{\partial \tau_i \partial \tau_j} \,\left(\partial^\mu \tau_i \partial_\mu \tau_j + \partial^\mu \theta_i \partial_\mu  \theta_j \right), \nn
\ee
where the leading order contribution to the kinetic Lagrangian for the $\tau$-fields is diagonalised by the transformation (\ref{volTrans}) where, as can be seen from (\ref{vol2}), $\chi$ corresponds to the heavy volume mode while $\phi$ is the light inflationary direction. Given that we are interested in just the inflaton coupling, we can keep $\chi$ at its classical value and expand only $\phi$ as $\phi=\langle\phi\rangle+\hat\phi$, obtaining from (\ref{volTrans}):
\be
\tau_1 = \langle\tau_1 \rangle\,e^{2k\hat\phi}\qquad\text{and}\qquad \tau_2=\langle\tau_2\rangle\,e^{-k\hat\phi}\,.
\ee
The kinetic Lagrangian for the axionic fields then gives:
\bea
\mc{L}_{\rm kin} &\supset& \frac{1}{4\tau_1^2}\, \partial_\mu \theta_1 \partial^\mu \theta_1 +  \frac{1}{2\tau_2^2}\,\partial_\mu \theta_2 \partial^{\mu} \theta_2 = \frac{e^{-4k\hat\phi}}{4\langle\tau_1\rangle^2}\, \partial_\mu \theta_1 \partial^\mu \theta_1 +  \frac{e^{2k\hat\phi}}{2\langle\tau_2\rangle^2}\,\partial_\mu \theta_2 \partial^{\mu} \theta_2 \nn \\
&\simeq& \frac{1}{4\langle\tau_1\rangle^2}\, \partial_\mu \theta_1 \partial^\mu \theta_1 \left(1-4k\hat\phi+8k^2\hat\phi^2+\dots\right) 
+  \frac{1}{2\langle\tau_2\rangle^2}\,\partial_\mu \theta_2 \partial^{\mu} \theta_2\left(1+2k\hat\phi + 2k\hat\phi^2+\dots\right). \nn
\eea
Hence the canonically normalised axions are given by:
\be
a_1 = \frac{\theta_1}{\sqrt{2}\langle\tau_1\rangle}  \qquad \text{and}\qquad a_2 =  \frac{\theta_2}{\langle\tau_2\rangle}\,,
\ee
and the Lagrangian describing the inflaton-axion-axion interactions takes the form:
\be
\mc{L}_{\hat\phi aa} = -\frac{2}{ \sqrt{3}}\, \frac{\hat\phi}{M_p}\partial_\mu a_1 \partial^\mu a_1 + 
\frac{1}{\sqrt{3}} \frac{ \hat\phi  }{M_p}\, \partial_{\mu} a_2 \partial^{\mu} a_2\,.
\ee
Using the equation of motion for $\hat\phi$, $\Box \hat\phi = m_\phi^2 \hat\phi$, and the fact that the axions are almost massless, i.e. $\Box a_1 \simeq \Box a_2 \simeq 0$, after integrating by parts and neglecting boundary terms, this interaction Lagrangian can be rewritten as: 
\be
\mc{L}_{\hat\phi aa} = \frac{1}{\sqrt{3}} \frac{m^2_\phi}{M_p} \,\hat\phi\, a_1 a_1 - \frac{1}{2\sqrt{3}}\frac{m^2_\phi}{M_p} \,\hat\phi\, a_2 a_2\,.
\label{Lintax}
\ee

\subsubsection{Visible sector fields}

The strength of the inflaton interaction with visible sector particles depends on the localisation of the MSSM (or generalisations thereof) in the extra dimensions.

\subsubsection*{Visible sector on D3-branes}

If the visible sector lives on D3-branes at singularities, the gauge kinetic function is set at tree-level by the dilaton, and so both the volume mode $\chi$ and the inflaton $\phi$ couple to visible sector gauge bosons only at loop level. All the other inflaton couplings to visible sector fields are further suppressed since \cite{Cicoli:2012aq,Higaki:2012ar,Angus:2014bia,Hebecker:2014gka,Cicoli:2015bpq}: ($i$) the couplings to quarks and leptons are chirality suppressed; ($ii$) depending on the level of sequestering of the visible sector from the sources of supersymmetry breaking, the decay of the inflaton to supersymmetric particles is either kinematically forbidden or mass suppressed; ($iii$) there is no tree-level inflaton coupling to Higgses induced by a Giudice-Masiero term in the K\"ahler potential since, as can be seen from (\ref{vol2}), the volume mode $\vo$ is given at leading order just by the heavy field $\chi$.\footnote{Ref. \cite{Cicoli:2010ha,Burgess:2010bz} showed that string loop corrections induce a subleading dependence of $\vo$ also on $\phi$ which however generates only a volume-suppressed inflaton coupling to Higgses.} Hence in the case of D3-branes, the main inflaton decay channel is into gauge bosons but the branching ratio into visible sector particles is negligible with respect to the one into hidden sector degrees of freedom since \cite{Angus:2012dd}: 
\be
\Gamma_{\hat\phi\rightarrow{\rm vis}}^\DT \simeq \left(\frac{\alpha_\SM}{4\pi}\right)^2 \Gamma_0\qquad\Rightarrow\qquad 
\frac{{\rm Br}\left(\hat\phi\rightarrow{\rm vis}\right)}{{\rm Br}\left(\hat\phi\rightarrow{\rm hid}\right)}\simeq \frac25 \left(\frac{\alpha_\SM}{4\pi}\right)^2\ll 1\,, \nn
\ee
where:
\be
\Gamma_0 \equiv \frac{1}{48\pi}\frac{m^3_\phi}{M_p^2}\,.
\label{G0}
\ee
Consequently, the D3-brane case is experimentally ruled out since it leads to a huge overproduction of axionic dark radiation with $\Delta N_{\rm eff}\gg 1$.

\subsubsection*{Visible sector on D7-branes}

We are therefore forced to consider models where the visible sector lives on D7-branes wrapped around the inflaton 4-cycle $D_1$ in order to maximise the inflaton branching ratio into SM fields. Let us stress that existing explicit global Calabi-Yau constructions of fibre inflation models feature always a stack of D7-branes wrapped around the fibre divisor in order to generate the open string loop corrections which develop the inflationary potential \cite{Cicoli:2016xae,Cicoli:2017axo}. Thus it is rather natural to identify this stack of D7-branes with the visible sector. Let us now analyse all possible inflaton decay channels into visible sector fields.

\begin{itemize}
\item \emph{Supersymmetric particles:} \\
In this case there is no sequestering of the MSSM from supersymmetry breaking, and so the soft terms are set by the gravitino mass, as in standard supergravity models: 
\be
M_{\rm soft} \simeq m_{3/2}\simeq 5\cdot 10^{15} \, \text{GeV}\,\gg m_\phi\simeq 5\cdot 10^{13} \, \text{GeV}\,.
\ee
Therefore the inflaton decay to supersymmetric particles is kinematically forbidden.\footnote{Rigorously speaking, $M_{\rm soft}$ is of order $m_{3/2}$ around the GUT scale and one should consider the RG running of the soft terms down to the energy scale $m_\phi \simeq 10^{13}$ GeV where the inflaton decays. However we can neglect this running since all the couplings involved are gravitational and the decay energy scale is relatively high.} 

\item \emph{Matter fermions:} \\
The inflaton decay into ordinary matter fermions like quarks and leptons with mass $m_f$ is chirality suppressed since: 
\be
\Gamma_{\hat\phi\rightarrow ff}\simeq \left(\frac{m_f}{m_\phi}\right)^2 \Gamma_0 \ll \Gamma_0\,.
\ee

\item \emph{Gauge bosons:} \\
The inflaton coupling to gauge bosons is induced by the moduli-dependence of the gauge kinetic function. In general we can have a stack of D7-branes wrapping the fibre divisor $D_1$ and another D7-stack wrapped around the base divisor $D_2$ with gauge kinetic functions respectively:
\be
f_1 = \frac{T_1}{2\pi}\qquad\text{and}\qquad f_2 =\frac{T_2}{2\pi}\,.
\label{fs}
\ee 
Given that the gauge couplings $g_i^{-2}$ are set by the real parts of the corresponding gauge kinetic functions, the base divisor $D_2$ can support just a hyper-weak gauge sector since from the minimisation relation (\ref{Anis}) we have:
\be
{\rm Re}(f_1) = \frac{1}{g_1^2} = \frac{\langle\tau_1\rangle}{2\pi} \simeq g_s^2 \frac{\langle\tau_2\rangle}{2\pi} 
\ll \frac{\langle\tau_2\rangle}{2\pi}={\rm Re}(f_2) = \frac{1}{g_2^2}\,,
\ee
which, for $g_s\lesssim 0.1$, implies $g_2^2\lesssim 0.01\,g_1^2$. As we have seen in Sec. \ref{SecInfl}, matching the observed amplitude of the density perturbations requires $\vo\simeq \sqrt{\tau_1}\tau_2\sim \mc{O}(10^3)$, which can be naturally realised for $\langle\tau_1\rangle \sim\mc{O}(10)$ and $\langle\tau_2\rangle\sim\mc{O}(10^3)$. Notice that the value of $\tau_1$ at the minimum both reproduces the right order of magnitude of SM gauge couplings and guarantees that the effective field theory is under control.  

We can then envisage two scenarios for the visible sector on the fibre divisor: ($i$) all D7-branes support a diagonal flux $F_1$ which makes the diagonal anomalous $U(1)$ heavy, leaving just a single $SU(5)$ or $SO(10)$ GUT-like gauge group; ($ii$) the D7-branes wrapping $D_1$ have different non-diagonal magnetic fluxes $F_{1,i}$ which break the gauge group to $SU(3)\times SU(2)\times U(1)$. In the first case there is just one gauge coupling given by the real part of $f_1$ in (\ref{fs}) shifted by the real part of the dilaton $s$ with an $F_1$-dependent coefficient $h(F_1)$, while in the second case there are three gauge couplings given by $2\pi g_i^{-2} = \tau_1 - h(F_{1,i}) s$. In both cases the inflaton interaction with SM gauge bosons can then be derived from the gauge kinetic Lagrangian:
\be
\mc{L}_{\rm kin}^{\rm gauge} = -\frac{1}{4 g^2} F^{\mu\nu} F_{\mu\nu} 
=  -\frac14 \hat{F}^{\mu\nu} \hat{F}_{\mu\nu} 
-\frac{\gamma}{2\sqrt{3}} \frac{\hat\phi}{M_p} \hat{F}^{\mu\nu} \hat{F}_{\mu\nu} +\dots\,,
\label{phiFF}
\ee
where the dots represent higher order interactions and the canonically normalised field strength looks like:
\be
\hat{F}^{\mu\nu} =  \sqrt{\frac{\langle\tau_1\rangle}{2\gamma\pi}} \,F^{\mu\nu}\qquad\text{with}\qquad \gamma\equiv \frac{\langle\tau_1\rangle}{\langle\tau_1\rangle - h \langle s\rangle}\,.
\label{gammadef}
\ee

\item \emph{Higgs bosons:} \\
Unsuppressed inflaton couplings to Higgs bosons can arise from a Giudice-Masiero term in the matter K\"ahler potential of the form:
\be
K_{\rm matter} \supset \tilde{K}_{H_u} \bar{H}_u H_u  +\tilde{K}_{H_d} \bar{H}_d H_d + Z \left(H_u H_d + {\rm h.c.}\right)\,.
\label{Kmatter}
\ee
The moduli-dependent functions $Z$, $\tilde{K}_{H_u}$ and $\tilde{K}_{H_d}$ are in general unknown and hard to compute since they are not holomorphic. Nonetheless, it is possible to infer their dependence on the $T$-moduli using the scaling behaviour of the physical Yukawa couplings \cite{Conlon:2006tj}. Moreover, as noticed in \cite{Angus:2012dd}, our Calabi-Yau volume $\vo \sim t_1 t_2^2$ has a structure similar to the volume form $\vo\sim t_1 t_2 t_3$ of simple toroidal orientifolds where these functions can be explicitly computed \cite{Ibanez:1998rf,Lust:2004cx,Lust:2004fi} (see also \cite{Aparicio:2008wh}). Depending on the microscopic origin of the Higgs bosons, there are three different possibilities for the K\"ahler matter metric (with $i=u,d$):
\be
\tilde{K}_{H_i,\parallel} = \frac{k_{H_i,\parallel}(U)}{\tau_2}\,, \qquad \tilde{K}_{H_i,\perp} = \frac{k_{H_i,\perp}(U)}{s}\,,
\qquad \tilde{K}_{H_i,\cap}=\frac{k_{H_i,\cap}(U)}{\sqrt{s \tau_2}}\,,
\label{three}
\ee
where the $k$'s are real functions of the complex structure moduli $U$ which can be considered as constants after the $U$-moduli are fixed at tree-level by background fluxes. The three matter metrics in (\ref{three}) correspond respectively to the cases where the Higgs doublets $H_u$ and $H_d$ are Wilson lines living on the D7-stack worldvolume counted by $h^{1,0}(D_i)$ with $i=1$ or $i=2$, deformation modes orthogonal to the D7-branes counted by $h^{2,0}(D_i)$ with $i=1$ or $i=2$, or open strings at the intersection between the two stacks of D7-branes wrapping the base and the fibre divisors. 

Explicit Calabi-Yau constructions of Fibre Inflation models feature a fibre divisor $D_1$ which is a K3 surface with $h^{1,0}(D_1)=0$ and $h^{2,0}(D_1)=1$, and a base 4-cycle $D_2$ that is a special deformation divisor with $h^{1,0}(D_2)=0$ and $h^{2,0}(D_2)\gg 1$ \cite{Cicoli:2016xae,Cicoli:2017axo}. Given that each Higgs doublet $H_u$ and $H_d$ needs more than just one complex degree of freedom, they can be only modes orthogonal to the base divisor. However, as pointed out above, this 4-cycle supports a hyper-weak sector. Moreover, as can be seen from (\ref{three}), in this case the matter K\"ahler metric would depend just on the dilaton, resulting in a leading order decoupling between the inflaton and the Higgses. The only option leftover is to realise the two Higgs doublets $H_u$ and $H_d$ via open strings at the intersection between two D7-stacks wrapping $D_1$ and $D_2$.\footnote{Another option might actually be to consider open strings stretched between differently magnetised D7-branes wrapping the K3 fibre $D_1$. However we do not discuss this situation since it gives the same result as the case with Higgs modes at the intersection between $D_1$ and $D_2$.} In what follows we shall therefore consider:
\be
\tilde{K}_{H_u} = \frac{k_{H_u}}{\sqrt{s \tau_2}} \qquad \text{and}\qquad  \tilde{K}_{H_d} = \frac{k_{H_d}}{\sqrt{s \tau_2}}\,, \nn
\ee
together with:
\be
Z =z\sqrt{\tilde{K}_{H_u} \tilde{K}_{H_d}}=\frac{z}{\left(s\tau_2\right)^{1/4}} \sqrt{k_{H_u} k_{H_d}} \,, \nn
\ee
where $z$ is real $U$-dependent parameter which can be treated just as an $\mc{O}(1)$ constant. If we now use (\ref{volTrans}) and expand $\phi$ around its minimum, the matter K\"ahler potential (\ref{Kmatter}) takes the form:
\bea
K_{\rm matter} &\supset& \frac{e^{\frac{k}{2}\hat\phi}}{\sqrt{s \langle\tau_2\rangle}} \left(k_{H_u}\bar{H}_u H_u  + k_{H_d} \bar{H}_d H_d \right)
+ \frac{z\sqrt{k_{H_u} k_{H_d}}}{\left(s\langle\tau_2\rangle\right)^{1/4}}\,e^{\frac{k}{4}\hat\phi}\left(H_u H_d + {\rm h.c.}\right) \nn \\
&=& \bar{\hat{H}}_u \hat{H}_u  + \bar{\hat{H}}_d \hat{H}_d + z \left(\hat{H}_u \hat{H}_d + {\rm h.c.}\right)  
\label{Kmexp} \\
&+& \frac{1}{2\sqrt{3}}\frac{\hat\phi}{M_p} \left(\bar{\hat{H}}_u \hat{H}_u  + \bar{\hat{H}}_d \hat{H}_d \right)
+ \frac{z}{4\sqrt{3}}\frac{\hat\phi}{M_p} \left(\hat{H}_u \hat{H}_d + {\rm h.c.}\right)+\dots, \nn
\eea
where the canonically normalised Higgs fields are given by:
\be
\hat{H}_i = \frac{\sqrt{k_{H_i}}}{\left(s \langle\tau_2\rangle\right)^{1/4}}\,  H_i \qquad \text{with}  \quad i= u,d\,.
\ee
As shown in \cite{Cicoli:2015bpq}, the first term in the third line of (\ref{Kmexp}) gives rise to inflaton decays into pairs of up-type or down-type Higgs fields but the corresponding decay rates are mass suppressed. On the other hand, the second term in the third line of (\ref{Kmexp}) generates an unsuppressed Giudice-Masiero interaction of the form:
\be
\mc{L}_\GM =  - \frac{z}{8\sqrt{3}}\frac{m^2_\phi}{M_p} \, \hat\phi \left( \hat{H}_u \hat{H}_d + \text{h.c.} \right), 
\label{LGM}
\ee
where:
\be
\hat{H}_u \hat{H}_d = \left(\hat{H}_u^+, \hat{H}_u^0\right) 
\begin{pmatrix}
0 & 1 \\
-1 & 0
\end{pmatrix}
\begin{pmatrix}
\hat{H}_d^0 \\
\hat{H}_d^-
\end{pmatrix}
= \hat{H}_u^+ \hat{H}_d^- - \hat{H}_u^0 \hat{H}_d^0  \,. \nn
\ee
Following \cite{Cicoli:2015bpq}, we now expand the Higgs complex fields as:
\bea
\hat{H}_u^+ &=& \frac{1}{\sqrt{2}} \left( h_1 + {\rm i}\, h_7 \right), \qquad 
\hat{H}_u^0 = \frac{1}{\sqrt{2}} \left( h_4 + {\rm i}\, h_6 \right), \nn \\
\hat{H}_d^- &=& \frac{1}{\sqrt{2}} \left( h_2 + {\rm i}\, h_8 \right), \qquad
\hat{H}_d^0 = \frac{1}{\sqrt{2}} \left( h_3+ {\rm i}\, h_5 \right). \nn
\eea
Hence the interaction Lagrangian (\ref{LGM}) can be rewritten as:
\be
\mc{L}_\GM = - \frac{z}{8\sqrt{3}}\frac{m^2_\phi}{M_p} \,  \hat\phi \left(h_1 h_2-h_3 h_4+h_5 h_6-h_7 h_8\right)\,.
\label{LGM2}
\ee
However in our case supersymmetry is broken at a high scale, and so only four Higgs degrees of freedom are light when the inflaton decays. In fact, electroweak symmetry breaking occurs at energies of order the gravitino mass while the inflaton decays at energies of order $m_\phi\ll m_{3/2}$. Thus we need to switch from the gauge eigenstates $h_i$ $i=1,..,8$ to the physical states $A^0$, $H^0$ and $H^\pm$, which acquire a mass of order $m_{3/2}$, the SM Higgs boson $h^0$ and the three would-be Goldstone bosons $G^0$ and $G^\pm$ via the following transformation: 
\bea
h_1 &=& \sin \beta\, {\rm Re} G^+ +  \cos \beta \, {\rm Re} H^+\,, \qquad \quad h_2 = -\cos \beta\, {\rm Re} G^+ + \sin \beta \, {\rm Re} H^+\,, \nn \\
h_3 &=&  \sqrt{2} v_d + \cos \alpha \,H^0 + \sin \alpha \, h^0\,, \qquad \,\,\,h_4 =  \sqrt{2} v_u  - \sin \alpha\, H^0 + \cos \alpha \, h^0\,, \nn \\
h_5 &=&  -\cos \beta \, G^0 +\sin \beta \, A^0\,, \qquad \qquad \quad h_6 = \sin \beta \, G^0 +\cos \beta \, A^0\,, \nn \\
h_7 &=& \sin \beta \, {\rm Im} G^+ + \cos \beta \, {\rm Im} H^+\,, \qquad \quad h_8 = \cos \beta \, {\rm Im} G^+ - \sin \beta \, {\rm Im} H^+\,, \nn
\eea
where $v_i= \langle \hat{H}_i^0\rangle$ with $i=u,d$, $\tan \beta = v_u/ v_d $ and: 
\be
\frac{\sin (2\alpha)}{\sin (2\beta)} = -\left( \frac{m^2_{H^0} + m^2_{h^0}}{m^2_{H^0} - m^2_{h^0}} \right)\simeq -1\,, \qquad \qquad 
\frac{\tan (2\alpha)}{\tan (2\beta)} = \left( \frac{m^2_{A^0} + m^2_Z}{m^2_{A^0} -  m^2_Z} \right)\simeq 1, \nn
\ee
which implies $\alpha\simeq \beta- \pi/2$. In order to get $m_{h^0}\simeq 125$ GeV in models with high scale supersymmetry breaking, one needs $\tan\beta\simeq 1$ \cite{Bagnaschi:2014rsa,Ellis:2017erg}, which fixes $\beta\simeq -\alpha \simeq \pi/4$. Therefore the terms in the Giudice-Masiero Lagrangian (\ref{LGM2}) which induce inflaton decays into light Higgs degrees of freedom are:
\bea
\mc{L}_\GM &\supset& \frac{z}{16\sqrt{3}}\frac{m^2_\phi}{M_p} \,\hat\phi 
\left[\sin(2 \beta) \, \left( |G^+|^2  + (G^0)^2  \right)  + \sin(2\alpha) \,  (h^0)^2\right] \nn \\
&\simeq& \frac{z}{16\sqrt{3}}\frac{m^2_\phi}{M_p} \,\hat\phi 
\left[|G^+|^2  + (G^0)^2  -  (h^0)^2 \right]\,.
\label{LGMfin}
\eea
\end{itemize}

\subsubsection{Dominant decay rates}

As described above, the leading inflaton decay channels are into: 
\begin{enumerate}
\item \emph{Hidden sector axions:} \\
The inflaton decay rate into ultra-light bulk axions arising from (\ref{Lintax}) turns out to be:
\be
\Gamma_{\hat\phi \rightarrow{\rm hid}} = \Gamma_{ \hat\phi \rightarrow a_1 a_1} + \Gamma_{ \hat\phi \rightarrow a_2 a_2} = \frac52\, \Gamma_0\,.
\ee

\item \emph{Visible sector gauge bosons:} \\
The inflaton decay rate into $N_g$ gauge bosons living on D7-branes wrapped around the fibre divisor is induced by the three-body interaction in (\ref{phiFF}) and looks like:
\be
\Gamma_{\hat\phi\rightarrow AA} = \gamma^2 N_g \Gamma_0\,.
\ee

\item \emph{Visible sector Higgses:} \\
The Giudice-Masiero Lagrangian (\ref{LGMfin}) yields the following inflaton decay rates into the SM Higgs and the three would-be Goldstone bosons:
\bea
\Gamma_{\hat\phi \rightarrow G^+ G^-} &=& \left( \frac{z}{16} \right)^2 \sin^2 (2\beta) \,\Gamma_0 \simeq \left( \frac{z}{16} \right)^2 \,\Gamma_0\,, \qquad
\Gamma_{\hat\phi \rightarrow G^0 G^0} = 2 \,\Gamma_{\hat\phi \rightarrow G^+ G^-}\,, \nn \\
\Gamma_{\hat\phi \rightarrow h^0 h^0} &=& \Gamma_{\hat\phi \rightarrow G^0 G^0} \left(\frac{\sin (2\alpha)}{\sin (2\beta)}\right)^2\simeq \Gamma_{\hat\phi \rightarrow G^0 G^0}\,. \nn
\eea
Thus the total inflaton decay rate into Higgs degrees of freedom becomes:
\be
\Gamma_{\hat\phi\rightarrow {\rm Higgses}}= \Gamma_{\hat\phi \rightarrow G^+ G^-} \left[3+2 \left(\frac{\sin (2\alpha)}{\sin (2\beta)}\right)^2\right]
\simeq 5\,\Gamma_{\hat\phi \rightarrow G^+ G^-}\simeq \left( \frac{z}{16} \right)^2 5 \,\Gamma_0\,. \nn
\ee
\end{enumerate}

We conclude that the total inflaton decay rate into visible and hidden sector fields is given, at leading order, by:
\be
\Gamma_{\hat\phi}^{\rm tot} = \Gamma_{\hat\phi\rightarrow {\rm vis}} + \Gamma_{\hat\phi\rightarrow {\rm hid}} =\left(c_{\rm vis}+c_{\rm hid}\right) \Gamma_0\,,
\ee
where:
\be
c_{\rm vis} = \gamma^2 N_g+ 5\left( \frac{z}{16} \right)^2\qquad\text{and}\qquad c_{\rm hid} = \frac52\,.
\label{coupl}
\ee
For $N_g=12$, the dependence of $c_{\rm vis}$ on $z$ is negligible for natural $\mc{O}(1)$ values of $\gamma$ and $z$ since:
\be
c_{\rm vis} = 12 \,\gamma^2 \left(1+ 1.6\cdot 10^{-3} \left( \frac{z}{\gamma} \right)^2\right)\simeq 12 \,\gamma^2 \,.
\ee

\subsection{Reheating temperature and number of efoldings}

After the end of inflation, for $\hat\phi < \hat\phi_{\rm ip}^{(1)}$, the inflaton start oscillating around its minimum behaving as non-relativistic matter with $w_{\rm rh}=0$. During this epoch, the inflaton energy density can be transferred to visible and hidden degrees of freedom via non-perturbative preheating due to a combination of tachyonic instability and parametric resonance, or via the standard perturbative decay of the inflaton condensate. Preheating for Fibre Inflation models has been studied in \cite{Antusch:2017flz} which showed that the field remains in practice homogeneous since none of the modes is significantly excited. Ref. \cite{Gu:2018akj} has instead recently found that inflaton particles can be produced due to parametric resonance even if not all of the energy of the inflaton condensate can be extracted via this non-perturbative process. The difference between the two results might be due to the fact that the analysis of \cite{Gu:2018akj} is based just on an analytical approximation while the study of \cite{Antusch:2017flz} is fully numerical, and so more reliable in a highly non-linear regime like preheating.\footnote{Ref. \cite{Barnaby:2009wr} studied preheating for a similar LVS inflationary model where the inflaton is a blow-up mode \cite{Conlon:2005jm}, and found a violent non-perturbative production of inflaton self-quanta due to the steepness of the inflaton potential around its minimum. However the final transfer of the inflaton energy to SM particles still occurs via perturbative decay. Notice that around the minimum the potential of Fibre Inflation models is shallower than the one of the inflationary model of \cite{Conlon:2005jm}, and so in our case preheating effects should be less important, in agreement with the results of \cite{Antusch:2017flz}.} Given the difficulty in Fibre Inflation to have non-perturbative preheating with an efficient production of inflaton particles, which have however to decay perturbatively later on to produce SM degrees of freedom, we expect that in our case the production of visible sector particles after the end of inflation is dominated by a purely perturbative dynamics. 

When the inflaton decays at energies of order $3 H \simeq 2 \Gamma_{\hat\phi}^{\rm tot}$, its energy density $\rho_{\rm tot} \simeq 3H^2 M_p^2$ gets transferred into visible and hidden sector particles with energies respectively:
\be
\rho_{\rm vis} = \frac{c_{\rm vis}}{c_{\rm tot}}\,\rho_{\rm tot}\qquad\text{and} \qquad \rho_{\rm hid}= \frac{c_{\rm hid}}{c_{\rm tot}}\,\rho_{\rm tot}\,,
\ee
where $c_{\rm tot}= c_{\rm vis}+ c_{\rm hid}$. Bulk axions interact only gravitationally \cite{Cicoli:2012aq,Higaki:2012ar,Angus:2014bia,Hebecker:2014gka,Cicoli:2015bpq}, and so they never reach thermal equilibrium, while visible sector degrees of freedom form a thermal bath with energy (in the approximation of sudden thermalisation):  
\be
\rho_{\rm vis} = \frac{\pi g_\ast(T_{\rm rh})}{30}\, T_{\rm rh}^4\,,
\ee
where $g_\ast(T_{\rm rh})$ is the number of relativistic degrees of freedom at reheating. Thus the reheating temperature is given by (restoring powers of $M_p$):
\be
T_{\rm rh} = \left( \frac{90}{\pi^2 g_\ast(T_{\rm rh})}\frac{c_{\rm vis}}{c_{\rm tot}} \right)^{1/4} \sqrt{H M_p} = 
\left( \frac{40 \,c_{\rm vis}\, c_{\rm tot}}{\pi^2 g_\ast(T_{\rm rh})}  \right)^{1/4} \sqrt{\Gamma_0 M_p}\,.
\label{TrhFI}
\ee
Using (\ref{G0}) and (\ref{coupl}) and setting $g_\ast (T_{\rm rh})= 106.75$, the reheating temperature can be rewritten in terms of the underlying parameters $m_\phi$, $\gamma$ and $z$ as:
\be
T_{\rm rh} \simeq  0.12 \,\gamma  
\left(1+\frac{5}{96\gamma^2}+8\cdot 10^{-4}\left( \frac{z}{\gamma} \right)^2\right)m_\phi\sqrt{\frac{m_\phi}{M_p}}\,,
\label{TrhFIfin}
\ee
where we took a Taylor expansion valid for $0\lesssim z\lesssim 10$. Setting $m_{\hat\phi}\simeq 5\cdot 10^{13}$ GeV from (\ref{massphi}), we obtain $T_{\rm rh}\simeq 3\,\gamma\cdot 10^{10}$ GeV. Notice that the order of magnitude of the reheating temperature is mainly set by the inflaton mass while it has a linear dependence on the underlying parameter $\gamma$ which is expected to be of $\mc{O}(1)$. Inserting $T_{\rm rh}\simeq 3\,\gamma\cdot 10^{10}$ GeV into the expression (\ref{Ne1}) for the number of efoldings, we end up with $N_e\simeq 52+\frac13 \ln\gamma\simeq 52$. 

\subsection{Dark radiation and tensor modes}

We can now use the value $N_e\simeq 52$ into the functions $n_s=n_s(N_e,R)$ and $r=r(N_e,R)$ obtained in Sec. \ref{SFI} to reduce them to $n_s=n_s(R)$ and $r=r(R)$. The scalar spectral index, and in turn the $g_s$-dependent parameter $R$, can be fixed from the requirement of matching Planck data. This can be done however only after specifying the value of $\Delta N_{\rm eff}$ to be used as a prior. 

The amount of extra dark radiation is determined by the ratio of the inflaton branching ratio into hidden and visible degrees of freedom \cite{Cicoli:2012aq,Higaki:2012ar}:
\be
\Delta N_{\rm eff} = \frac{43}{7} \frac{\Gamma_{\hat\phi\rightarrow {\rm hid}}}{\Gamma_{\hat\phi\rightarrow {\rm vis}}} 
\left( \frac{g_\ast (T_{\rm dec})}{g_\ast (T_{\rm rh})} \right)^{1/3} 
\simeq \frac{0.6}{\gamma^2}\left(1+1.6\cdot 10^{-3}\left(\frac{z}{\gamma}\right)^2\right)^{-1}\simeq \frac{0.6}{\gamma^2}\,, 
\label{NeffPred}
\ee
where the number of relativistic species at neutrino decoupling and at reheating are respectively $g_\ast (T_{\rm dec})=10.75$ and $g_\ast (T_{\rm rh}) =106.75$. This prediction depends crucially on the exact value of the parameter $\gamma$ defined in (\ref{gammadef}). Its microscopic expression is given in terms of the function $h(F_1)$ which depends on the gauge flux $F_1$ on the brane wrapping the fibre divisor $D_1$. If this flux is expanded in a basis of $(1,1)$-forms $\hat{D}_i$, $i=1,2,3$, as $F_1=2\pi n_i \hat{D}_i$ with $n_i \in \mathbb{Z}$, the function $h(F_1)$ takes the form \cite{Jockers:2005zy,Haack:2006cy,Cicoli:2011yh}:
\be
h(F_1) = \frac12\,k_{1ij} \,n_i\, n_j = \frac12\,k_{122}\, n_2^2 \geq 0\,,
\ee
since, as can be seen from the volume form (\ref{volts}), for K3-fibred Calabi-Yau threefolds the only non-vanishing intersection number of the form $k_{1ij}$ is $k_{122}$. Notice that $\gamma$ depends also on the value of the fibre modulus at the minimum $\langle\tau_1\rangle$ which is given in (\ref{tau1soln2}) in terms of the underlying quantities. Hence $\gamma$ depends on $6$ microscopic parameters, $n_2$, $k_{122}$, $g_s$, $c_1^\KK$, $c^\W$ and $\langle\vo\rangle$, which can be reduced to just $3$ by setting the coefficients of the string loop corrections to natural $\mc{O}(1)$ values, $c_1^\KK=c^\W=1$, and $\langle \vo\rangle=10^3$ to reproduce the correct observed amplitude of the density perturbations as explained below (\ref{Minf}). Moreover from (\ref{fs}) we realise that $\alpha_{\rm vis}^{-1} = 2\left(\langle\tau_1\rangle-h(F_1)\, g_s^{-1}\right)$ which implies:
\be
\gamma = 2\,\alpha_{\rm vis}\,\langle\tau_1\rangle = 1 + 2\alpha_{\rm vis}\,\frac{h(F_1)}{g_s} \geq 1\,,
\label{gam}
\ee
since $h(F_1)$ is positive semi-definite. From (\ref{NeffPred}) this observation sets an important upper bound on the prediction for dark radiation: $\Delta N_{\rm eff}\lesssim 0.6$. 

Notice that $n_2=0$ implies $\gamma=1$, $\alpha_{\rm vis}^{-1}=2\langle\tau_1\rangle$ and $\Delta N_{\rm eff}\simeq 0.6$ regardless of the values of $g_s$ and $k_{122}$. An additional phenomenological condition which can be imposed to fix the string coupling for any value of $k_{122}$ using (\ref{tau1soln2}) is $\alpha_{\rm vis}^{-1}=25$ which implies $\langle \tau_1\rangle =12.5$. In order to consider this case as phenomenologically viable, we have however to check that a vanishing flux integer $n_2$ is still compatible with the presence of chiral matter on the D7-stack wrapping $D_1$. This is indeed the case since the number of chiral zero-modes at the intersection between the D7-branes wrapping $D_1$ with gauge flux $F_1$ and $D_2$ with gauge flux $F_2=2\pi m_i \hat{D}_i$ with $m_i \in \mathbb{Z}$ is:
\be
I_{{\rm D7}_1 {\rm D7}_2} = \frac{1}{2\pi}\int_X \left(F_1-F_2\right)\wedge \hat{D}_1 \wedge \hat{D}_2 = k_{122} \left(n_2-m_2\right)\,,
\ee
which for $n_2=0$ can still be non-zero if $m_2\neq 0$. Notice finally that the cancellation of Freed-Witten anomalies forces in general the presence of non-zero half-integer fluxes \cite{Minasian:1997mm,Freed:1999vc}. However this is not the case for non-spin 4-cycles like the K3 divisor $D_1$ \cite{Cicoli:2011qg}. This implies that $n_2=0$ is a consistent choice. 

If we now consider the case with $n_2\neq 0$, $\gamma$ becomes larger than $1$, and so $\Delta N_{\rm eff}$ quickly decreases below $0.6$. However in this case the phenomenological requirement $\alpha_{\rm vis}^{-1}= 25$ does not fix $\langle\tau_1\rangle$ which tends however to become larger than in the $n_2=0$ case since from (\ref{gam}) it can be written as $\langle\tau_1\rangle =  12.5\,\gamma\geq 12.5$. Hence larger values of $\langle\tau_1\rangle$ correspond to larger values of $\gamma$ which correlate with smaller values of $\Delta N_{\rm eff}$.

\begin{figure}[!htbp]
\centering
\includegraphics[scale = 0.7]{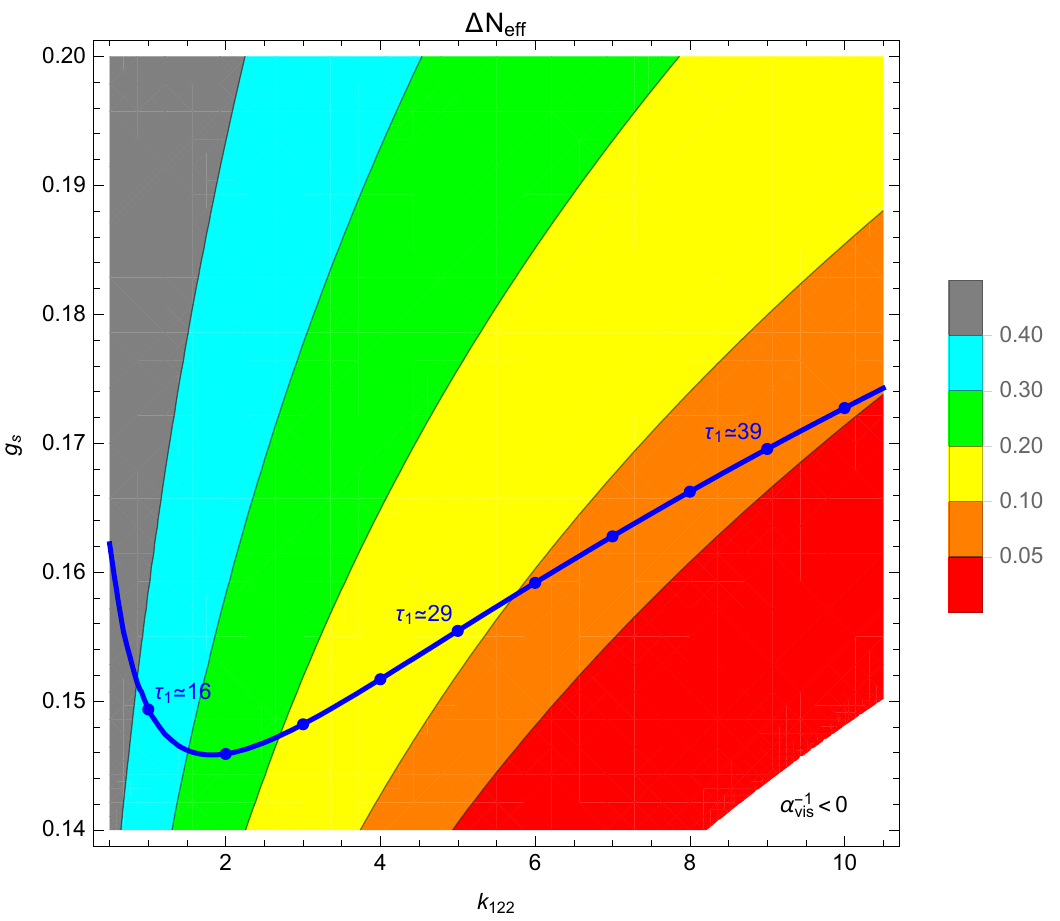}
\includegraphics[scale = 0.7]{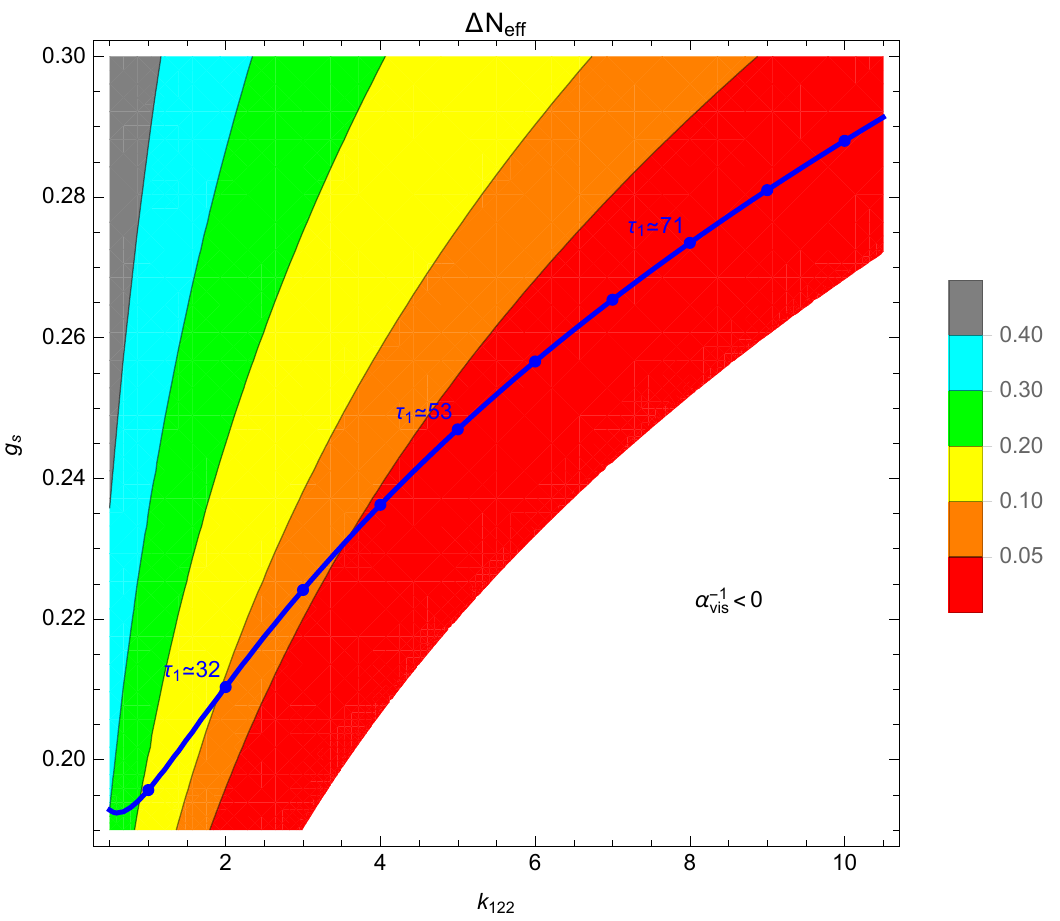}
\caption{Dark radiation predictions for $n_2=1$ on the left and $n_2=2$ on the right. The volume has been fixed at $\langle\vo\rangle= 10^3$ and the blue lines correspond to $\alpha_{\rm vis}^{-1}=25$. We have also indicated different values of $\langle\tau_1\rangle$ at fixed $\alpha_{\rm vis}^{-1}=25$.}
\label{Figure3}
\end{figure}

\begin{figure}[!htbp]
\centering
\includegraphics[scale = 0.7]{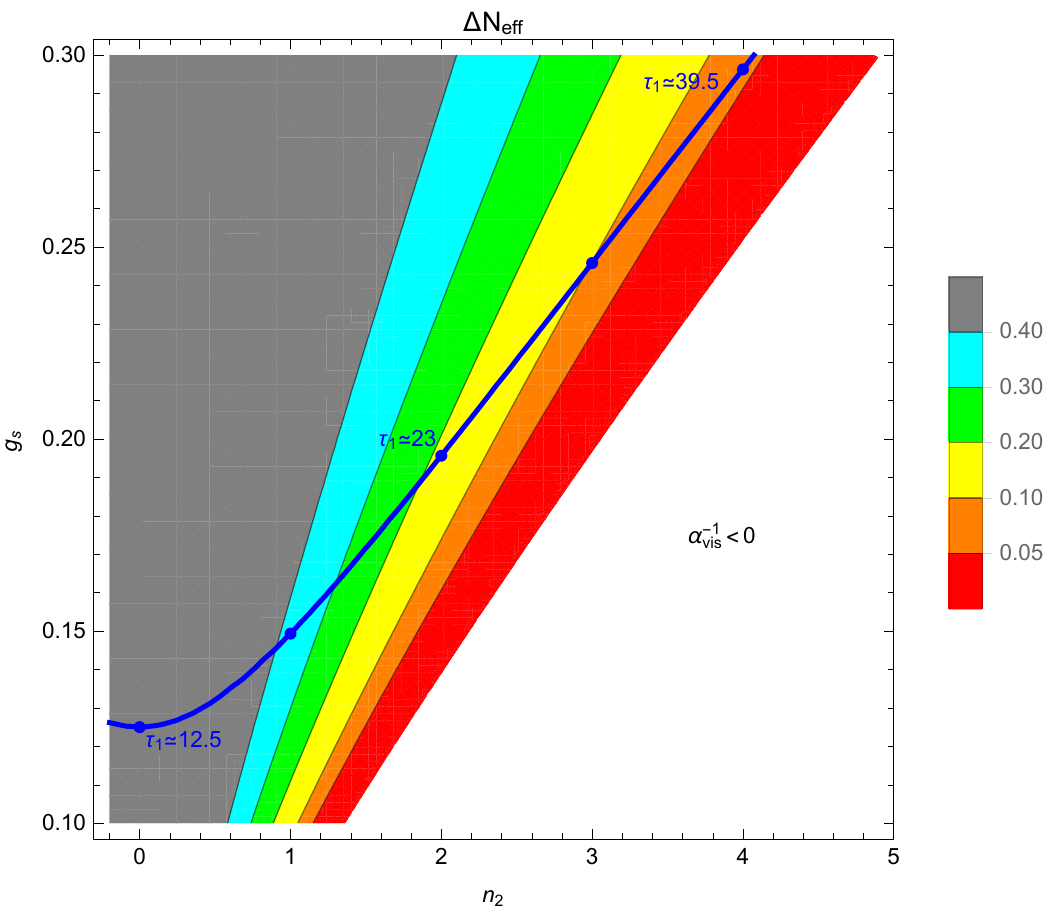}
\includegraphics[scale = 0.7]{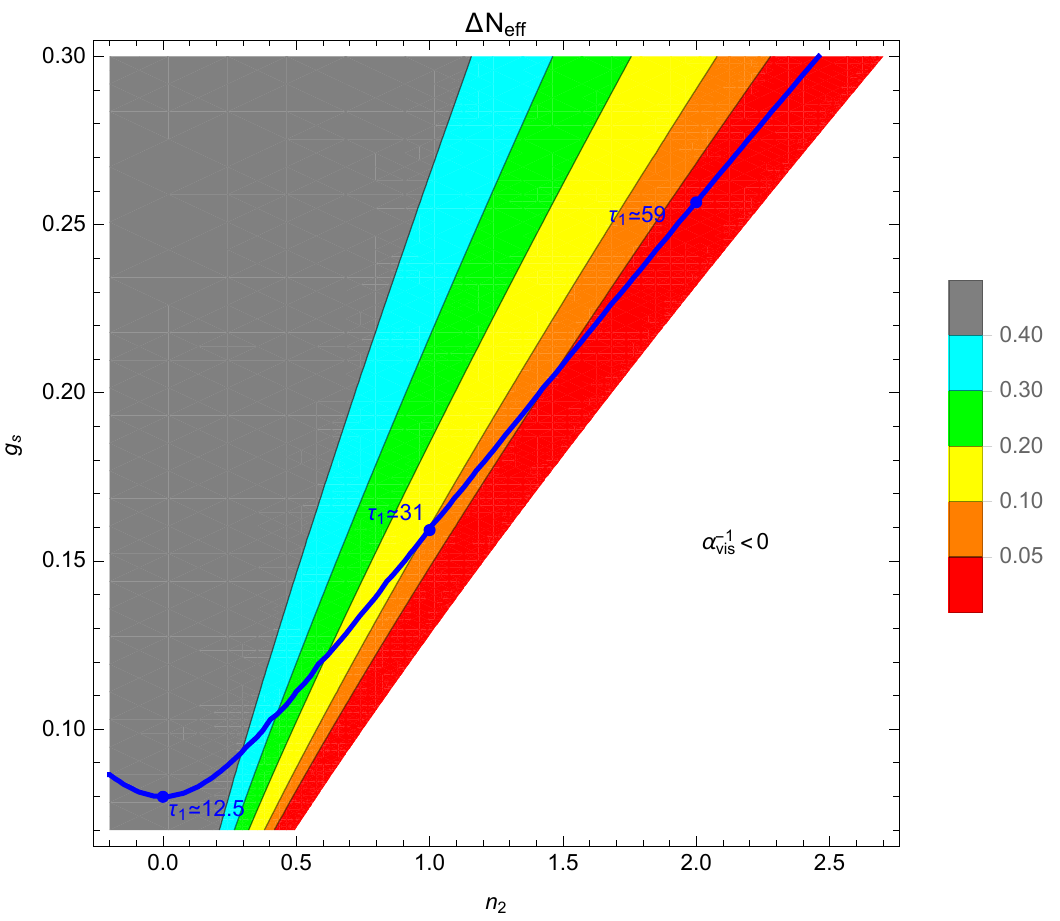}
\caption{Dark radiation predictions for $k_{122}=1$ on the left and $k_{122}=6$ on the right. The volume has been fixed at $\langle\vo\rangle= 10^3$ and the blue lines correspond to $\alpha_{\rm vis}^{-1}=25$. We have also indicated different values of $\langle\tau_1\rangle$ at fixed $\alpha_{\rm vis}^{-1}=25$.}
\label{Figure4}
\end{figure}

The prediction for $\Delta N_{\rm eff}$ as a function of the string coupling $g_s$ are shown in Fig. \ref{Figure3} and \ref{Figure4} for values $g_s\lesssim 0.3$ where perturbation theory is still under control. In Fig. \ref{Figure3} the intersection number $k_{122}$ varies while the flux integer $n_2$ is fixed at $n_2=1$ and $n_2=2$, whereas in Fig. \ref{Figure4} $n_2$ varies while $k_{122}$ is kept fixed at $k_{122}=1$ and $k_{122}=6$. The blue lines represent the condition $\alpha_{\rm vis}^{-1}=25$ and the value of $\langle\tau_1\rangle$ is shown explicitly for some reference values along these lines. 

This analysis shows that there are two qualitatively different regimes:
\bi
\item \textbf{Fluxed D7 on the fibre divisor and small extra dark radiation}\\
If $n_2\neq 0$, $\Delta N_{\rm eff}\lesssim 0.2$ except for $n_2=1$ and $k_{112}\leq 2$. Due to the small value of $\Delta N_{\rm eff}$, this case is in perfect agreement with Planck observations which favour a scalar spectral index centered around $n_s\simeq 0.965$ \cite{Aghanim:2018eyx}. As explained in Sec. \ref{SFI}, these data can be reproduced in the inflationary case $1$ where the parameter $R$ takes a value which is small enough to have horizon exit in a region where the positive exponential can be neglected. As can seen from Fig. \ref{Fig1b}, \ref{Fig2b} and \ref{Fig3}, this can be achieved if the parameter $R\ll R_0=2.7\cdot 10^{-5}$ . Let us make three important observations:
\ben
\item There is no fine-tuning associated with this small value of $R$ since, as can be seen from (\ref{V0}), this parameter scales as $R\propto g_s^4\ll 1$. 

\item The prediction for primordial tensor modes is around $r\simeq 0.007$.

\item As can be seen from  Fig. \ref{Figure3} and \ref{Figure4}, $\Delta N_{\rm eff}\lesssim 0.2$ correlates with $\langle\tau_1\rangle\gtrsim 20$. Large values of the fibre modulus at the minimum at fixed overall volume $\langle\vo\rangle=10^3$ might be in tension with the requirement of obtaining enough efoldings of inflation since, as shown in general \cite{Cicoli:2018tcq} and as found in particular explicit Calabi-Yau examples in \cite{Cicoli:2016xae,Cicoli:2017axo}, the inflaton field range is upper bounded due to the K\"ahler cone conditions. However, in case $1$ this is not a problem since the potential at large field values is very flat, and so $N_e\simeq 52$ can be easily obtained even in the presence of a large value of the inflaton at the minimum and an inflaton upper bound. 
\een
Let us present an explicit example which reproduces this case. The underlying parameters are:
\be
c_1^\KK=1\quad c_2^\KK = 0.1\quad c^\W = 10\quad g_s = 0.173 \quad W_0=65\quad \langle\vo\rangle=10^3\quad k_{122}=10\quad n_2=1\,,  \nn
\ee
which give:
\be
R = 1.78\cdot 10^{-7}\qquad \langle\tau_1\rangle = 41.45\qquad\alpha_{\rm vis}^{-1}=25\qquad \Delta N_{\rm eff} \simeq 0.05\,.
\ee
Inflation ends at $\hat\phi_{\rm end} = 0.917$ where $\epsilon(\hat\phi_{\rm end})=1$ while horizon exit takes place at $\hat{\phi}_* = 5.801$ where $N_e(\hat\phi_*)=52$. The observed amplitude of the scalar fluctuations is correctly matched and the prediction for the two main cosmological observables is:
\be
n_s=  0.965 \qquad\text{and}\qquad r=0.0065\,. 
\ee
Notice that the total field range including the inflaton minimum is given by $\hat\phi_*$ which can very well be below the geometrical upper bound computed in \cite{Cicoli:2018tcq}:
\be
\hat\phi_*=5.801< \Delta \phi = \frac{\sqrt{3}}{2}\ln\left(\frac{b \langle\vo\rangle}{a\langle\tau_s\rangle^{3/2}}\right) = 
5.982 +\frac{\sqrt{3}}{2}\ln\left(\frac{b}{a\langle\tau_s\rangle^{3/2}}\right)\,,
\ee
if the two parameters $a$ and $b$ which depend on the Calabi-Yau topology are such that $b\gtrsim a \langle\tau_s\rangle^{3/2}$.

\item \textbf{Fluxless D7 on the fibre divisor and large extra dark radiation}\\
If the gauge flux integer $n_2$ is turned off, the prediction for the abundance of extra dark radiation is $\Delta N_{\rm eff}\simeq 0.6$. As soon as $n_2\neq0$, $\Delta N_{\rm eff}$ quickly goes below $0.2$ unless $n_2=k_{112}=1$ which gives $\Delta N_{\rm eff}\simeq 0.37$, or $n=1$ and $k_{112}=2$ which yields $\Delta N_{\rm eff}\simeq 0.25$. Due to the large value of $\Delta N_{\rm eff}$, we have to reproduce a scalar spectral index centered around $n_s\simeq 0.99$ \cite{Ade:2015xua}. As shown in Fig. \ref{Fig1b}, \ref{Fig2b} and \ref{Fig3}, we need therefore to focus on the inflationary case $2$. In this case the potential in the horizon exit region is steeper due to the effect of the positive exponential term, and so the tensor-to-scalar ratio is slightly higher: $r\simeq 0.01$. The steepness of the potential tends also to increase the field range needed to achieve enough efoldings of inflation. However, given that $\langle\tau_1\rangle\lesssim 20$, we expect also in this case to be able to achieve $N_e\simeq 52$ without violating any geometrical inflaton upper bound from the K\"ahler cone conditions. 

An explicit example which reproduces this case features the following microscopic parameters:
\be
c_1^\KK= c_2^\KK = 1\quad c^\W = 6\quad g_s = 0.149 \quad W_0=60\quad \langle\vo\rangle=10^3\quad k_{122}=1\quad n_2=1\,,  \nn
\ee
which give:
\be
R = 2.76\cdot 10^{-5}\qquad \langle\tau_1\rangle = 15.85\qquad\alpha_{\rm vis}^{-1}=25\qquad \Delta N_{\rm eff} \simeq 0.37\,.
\ee
Inflation ends at $\hat\phi_{\rm end} = 0.918$ where $\epsilon(\hat\phi_{\rm end})=1$ while horizon exit takes place at $\hat{\phi}_* = 5.945$ where $N_e(\hat\phi_*)=52$. The observed amplitude of the scalar fluctuations is correctly matched and the prediction for the two main cosmological observables is:
\be
n_s=  0.99 \qquad\text{and}\qquad r=0.01\,. 
\ee
Notice again that the total field range including the inflaton minimum is given by $\hat\phi_*=5.945$ which can very well be below the geometrical upper bound computed in \cite{Cicoli:2018tcq}:
\be
\hat\phi_*=5.945< \Delta \phi = \frac{\sqrt{3}}{2}\ln\left(\frac{b \langle\vo\rangle}{a\langle\tau_s\rangle^{3/2}}\right) = 
5.982 +\frac{\sqrt{3}}{2}\ln\left(\frac{b}{a\langle\tau_s\rangle^{3/2}}\right)\,,
\ee
if the topology-dependent $a$ and $b$ parameters are such that $b\gtrsim a \langle\tau_s\rangle^{3/2}$.
\ei

\section{Conclusions}
\label{SecConcl}

Fibre Inflation is a very promising string framework to build inflationary models \cite{Cicoli:2008gp, Burgess:2016owb,Broy:2015zba,Cicoli:2016chb} since it leads to natural inflaton candidates protected by an approximate non-compact shift symmetry \cite{Burgess:2014tja} and allows one to construct explicit Calabi-Yau embeddings with chiral matter \cite{Cicoli:2016xae, Cicoli:2017axo}. However, in order to make clear predictions for the two main cosmological observables, the scalar spectral index $n_s$ and the tensor-to-scalar ratio $r$, one has to study the post-inflationary evolution of this class of models.

A crucial quantity that has to be determined is the reheating temperature $T_{\rm rh}$ which fixes the number of efoldings $N_e$ between horizon exit and the end of inflation. In this paper we studied perturbative reheating in Fibre Inflation models and computed the inflaton decay rates into visible and hidden sector particles. The reheating temperature turns out to be of order $T_{\rm rh}\simeq 10^{10}$ GeV which, when inserted into (\ref{Ne1}), gives an exact number of e-foldings $N_e\simeq 52$. 

A typical feature of string compactifications, which cannot be neglected in the study of post-inflationary string cosmology, is the generic presence of ultra-light axions \cite{Svrcek:2006yi, Conlon:2006tq, Arvanitaki:2009fg, Cicoli:2012sz} which behave as extra relativistic degrees of freedom and give rise to non-zero contributions to $\Delta N_{\rm eff}$ \cite{Cicoli:2012aq,Higaki:2012ar,Angus:2014bia,Hebecker:2014gka,Cicoli:2015bpq}. Fibre Inflation models feature two ultra-light bulk axions with a direct coupling to the inflaton. If the visible sector is realised on a stack of D7-branes wrapped around the inflaton divisor, we found that the main inflaton decay channels are into Higgses, gauge bosons and bulk axions. 

The prediction for $\Delta N_{\rm eff}$ depends crucially on the value of the gauge flux on the visible sector D7-stack. If this flux is non-zero, the inflaton coupling to visible gauge bosons is effectively enhanced, and so $\Delta N_{\rm eff}$ turns out to be almost negligible, in perfect agreement with observations. When used as a prior to interpret Planck data, an almost zero value of $\Delta N_{\rm eff}$ requires $n_s\simeq 0.965$ \cite{Aghanim:2018eyx}
. We then used the function $n_s=n_s(N_e,R)$ obtained in Sec. \ref{SFI} to fix $R$ from the knowledge of $n_s$ and $N_e$. We found that $R$ has to be chosen such that $R\ll R_0\equiv 2.7\cdot 10^{-5}$, which can be obtained naturally since $R\propto g_s^4 \ll 1$. The relation $r=r(N_e,R)$ gave us the final prediction $r\simeq 0.007$. 

On the other hand, if the visible sector D7-stack does not support any gauge flux, the branching ratio for the inflaton decay into ultra-light axions is larger, leading to $\Delta N_{\rm eff}\lesssim 0.6$. Such a large amount of extra dark radiation, when used as a prior for Planck data, requires $n_s\simeq 0.99$ \cite{Ade:2015xua}. In turn, the coefficient which controls the strength of positive exponential contributions to the inflationary potential has to increase, $R\gtrsim R_0$, giving rise to a larger tensor-to-scalar ratio of order $r\simeq 0.01$.

Notice that a direct inflaton coupling to Higgs degrees of freedom is induced by a Giudice-Masiero coupling in the matter K\"ahler potential \cite{Cicoli:2012aq,Higaki:2012ar}. This coupling was crucial to avoid dark radiation overproduction in K\"ahler Moduli Inflation where the inflaton is a blow-up mode \cite{Conlon:2005jm}. In this inflationary scenario, reheating is driven by the decay of the lightest modulus \cite{Cicoli:2012aq,Higaki:2012ar,Cicoli:2016olq} and the visible sector is localised on D3-branes at singularities \cite{Conlon:2008wa,Cicoli:2012vw}. However, in Fibre Inflation models, we found that the prediction for $\Delta N_{\rm eff}$ is almost insensitive to the value of the Giudice-Masiero coupling. This is an advantage of these models since the exact form of the Giudice-Masiero term in the K\"ahler potential is still only poorly understood. 

Moreover, it is important to stress that the generic case features non-zero gauge fluxes on the D7-branes wrapping the inflaton divisor. Hence Fibre Inflation models are naturally not affected by any dark radiation overproduction problem even if they are characterised by the presence of two ultra-light axion degrees of freedom. As pointed out in \cite{Cicoli:2018ccr}, in a curved field manifold, these fields might have an important effect on the inflationary dynamics which has still to be studied in detail.\footnote{Ref. \cite{Gu:2018akj} has shown that there is no parametric amplification of the volume entropy mode at preheating. This is an expected result due to the large mass of the volume mode. This is instead not necessarily the case for the two bulk axions since they are almost massless.} After inflation, these ultra-light axions can have several interesting cosmological applications like curvaton modes \cite{Burgess:2010bz}, particles responsible for the origin of astrophysical lines \cite{Cicoli:2014bfa, Cicoli:2017zbx}, fuzzy dark matter \cite{Hui:2016ltb} or quintessence fields \cite{Cicoli:2018kdo}. 

We finally point out that the case with large $\Delta N_{\rm eff}$ might be useful for one theoretical and two phenomenological reasons:
\ben
\item As we have seen, $n_s\simeq 0.965$ can be obtained only when $R\ll R_0$. In some models this might not be possible from the theoretical point of view. All the coefficients appearing in the parameter $R$ defined in (\ref{V0}) are tunable, and so in a flux landscape one should always have enough tuning freedom to make $R$ small enough. However, on top of string loops, also higher derivative $\alpha'$ terms can generate positive exponential contributions to the inflationary potential \cite{Ciupke:2015msa} whose coefficient might not have enough tuning freedom to realise $R\ll R_0$ \cite{Grimm:2017okk}. Even if these $F^4$ terms have still to be properly understood, in the case where they forbid the $R\ll R_0$ regime, one would need to set $n_2=0$ which gives $\Delta N_{\rm eff}\lesssim 0.6$ and $n_s\simeq 0.99$. As we have shown, this value of $n_s$ can be achieved for larger value of $R$ of order $R\gtrsim R_0$. However in this case the potential would be steeper for large field values, and so one would have to check that enough efoldings of inflation can be achieved before hitting the walls of the K\"ahler cone \cite{Cicoli:2018tcq}.

\item From the phenomenological point of view, the next generation of cosmological observations might be sensitive to values of the tensor-to-scalar ratio as small as $r\simeq 0.01$ \cite{Cabass:2015jwe}. Such a value of $r$ would still be too large for the case with negligible $\Delta N_{\rm eff}$. Hence a detection of primordial gravity waves with $r\simeq 0.01$ can be accommodated in Fibre Inflation models only for $\Delta N_{\rm eff}\lesssim 0.6$.

\item A large contribution of extra dark radiation of order $\Delta N_{\rm eff}\lesssim 0.6$ might be useful to solve the existing tension between CMB and low-redshift measurements of the present Hubble constant $H_0$ \cite{Riess:2016jrr}. In fact, assuming a standard $\Lambda$CDM scenario, Planck 2018 data give $H_0 = 67.27\pm 0.60$ km s$^{-1}$ Mpc$^{-1}$ \cite{Aghanim:2018eyx}, which is in $3.6\sigma$ tension with direct supernovae measurements which give $H_0 = 73.52\pm 1.62$ km s$^{-1}$ Mpc$^{-1}$ \cite{Riess:2018byc}. However, if $\Delta N_{\rm eff}=0.39$ is used as a prior to interpret CMB observations, Planck 2015 data give a larger value of the Hubble constant, $H_0 = 70.6\pm 1.0$ km s$^{-1}$ Mpc$^{-1}$ \cite{Ade:2015xua}. Moreover, the combined Planck+galaxy BAO+HST analysis of \cite{Aghanim:2018eyx} gives $N_{\rm eff} = 3.27 \pm 0.15$ and $H_0 = 69.32\pm 0.97$ km s$^{-1}$ Mpc$^{-1}$, while the combination of Planck+galaxy BAO+LyaF BAO measurements yields $N_{\rm eff} = 3.43 \pm 0.26$ and $H_0 = 71 \pm 1.7$ km s$^{-1}$ Mpc$^{-1}$ \cite{Aubourg:2014yra}. Finally the comprehensive analysis of \cite{Riess:2016jrr}, which includes Planck+galaxy BAO+LyaF BAO+HST+SN data, has found $N_{\rm eff} = 3.41 \pm 0.22$ and $H_0 = 70.4\pm 1.2$ km s$^{-1}$ Mpc$^{-1}$. Notice however that a non-zero $\Delta N_{\rm eff}$ would increase the tension in $\sigma_8$ between Planck data and measurements of the growth rate of the density perturbations. It is therefore not entirely clear yet if the solution of this $H_0$ tension relies indeed on the presence of extra dark radiation. 
\een
Let us finally mention that our construction of the visible sector is chiral but we did not discuss how to obtain matter in the right representations to realise the Standard Model or a GUT theory. This is beyond the scope of our paper which is already representing one of the few examples of a string model where one can describe inflation and reheating in the presence of a chiral visible sector. However an exact realisation of the Standard Model would be necessary to fully trust our results. We leave this investigation for the future.

\section{Acknowledgements}

We would like to thank D. Ciupke, A. Di Marco, F. Finelli, F. Muia, G. Pradisi and F. Quevedo for useful discussions.

\end{document}